\documentclass[times,english,sort&compress]{elsarticle}
\pdfoutput=1
\usepackage{color}
\usepackage[colorlinks=true] {hyperref}
\usepackage{float}
\usepackage{stmaryrd}
\usepackage{babel}
\usepackage{units}
\usepackage{textcomp}
\usepackage{textcomp}
\usepackage{amsmath}
\usepackage{commath}
\usepackage{amssymb}
\usepackage{graphicx}
\usepackage{esint}

\usepackage{geometry}
\makeatletter

\usepackage{epstopdf}

\journal{Communications in Nonlinear Science and Numerical Simulation}

\begin{document}
\begin{frontmatter}

\title{Two-dimensional composite solitons in a spin-orbit-coupled Fermi gas in free space}

\author[UFRO]{Pablo D\'{\i}az}
\author[UTA]{David Laroze\corref{mycorrespondingauthor}}
\cortext[mycorrespondingauthor]{Corresponding author}
\ead{dlarozen@uta.cl}
\author[UFRO2,Kassel]{Andr\'{e}s \'{A}vila}
\author[TELAVIV]{Boris A. Malomed}

\address[UFRO]{Departamento de Ciencias F\'{i}sicas, Universidad de La Frontera, Casilla 54-D, Temuco, Chile.}
\address[UTA]{Instituto de Alta Investigaci\'{o}n, CEDENNA, Universidad de Tarapac\'{a}, Casilla 7D, Arica, Chile.}
\address[UFRO2]{Departamento de Ingenier\'{i}a Matem\'{a}tica, Centro de Excelencia de Modelaci\'on y Computaci\'on Cient\'{\i}fica, Universidad de La Frontera, Casilla 54-D, Temuco, Chile.}
\address[Kassel]{Institut f\"ur Mathematik, Universit\"at Kassel,  34132 Kassel, Germany}
\address[TELAVIV]{Department of Physical Electronics, School of Electrical Engineering, Faculty of Engineering, and Center for Light-Matter Interaction, Tel Aviv University, Tel Aviv 69978, Israel.}

\begin{abstract}
We address a possibility of creating soliton states in oblate
binary-fermionic clouds in the framework of the density-functional theory,
which includes the spin-orbit coupling (SOC) and nonlinear attraction
between spin-up and down-polarized components of the spinor wave function.
In the limit when the inter-component attraction is much stronger than the
effective intra-component Pauli repulsion, the resulting model also
represents a system of Gross-Pitaevskii equations for a binary Bose-Einstein
condensate including the SOC effect. We show that the model gives rise to
two-dimensional quiescent composite solitons in free space. A stability
region is identified for solitons of the mixed-mode type (which feature
mixtures of zero-vorticity and vortical terms in both components), while
solitons of the other type, semi-vortices (with the vorticity carried by one
component) are unstable. Due to breaking of the Galilean invariance by SOC,
the system supports moving solitons with velocities up to a specific
critical value. Collisions between moving solitons are briefly considered
too. The collisions lead, in particular, to a quasi-elastic rebound, or an
inelastic outcome, which features partial merger of the solitons.
\end{abstract}

\begin{keyword}
Cold Atoms\sep spin-orbit coupling\sep solitons.
\end{keyword}

\end{frontmatter}

\section{Introduction}
\label{sec-1}

Experimental and theoretical research of degenerate fermionic gases has
drawn much interest, as their confining geometry, size, and interaction
strength can be varied in broad ranges \cite%
{Pit08,Blo08,Baumann,Zipkes,Perron}.With the help of many-body Hamiltonian
models, phase diagrams of quasi-one-dimensional (1D) Fermi systems were
produced in several settings \cite%
{Lieb63a,Lieb63b,Gaudin67,Yang67,Fuchs04,Liu07,Giraud09,Blume09}. The
mean-field theory, which essentially amounts to the establishment of the
Gross Pitaevskii equations (GPEs), provides a very accurate dynamical model
for bosonic gases \cite{Pit}. Models for Fermi superfluids and Fermi-Bose
mixtures were elaborated, under specific conditions, in the form of the
density-functional theory \cite{Salasnich00a}-\cite{Pu}. Using these
theoretical models, various nonlinear effects in degenerate quantum gases
were studied, such as generation of different types of vortex rings \cite%
{Anderson01,Ginsberg05,Shomroni09}, shock waves \cite{Dutton01,Chang08}, and
chaotic dynamics \cite{Ott03}. Other nonlinear effects in degenerate Fermi
systems, as well as in Bose-Fermi mixtures, were elaborated in Refs. \cite%
{Yefsah2007,Antezza2007,Adhikari2008,Diaz2012,Yefsah2013,Jie2014,Diaz2015,Lombardi2017,Shang2018}.

Creation of stable 2D and 3D solitons, both fundamental and vortical ones,
is a well-known fundamental problem in nonlinear optics and Bose-Einstein
condensates (BECs). Induced by the Kerr effect in optics and attractive
inter-atomic interactions in BEC, the basic cubic self-focusing nonlinearity
builds solitons which are completely unstable because of the critical and
supercritical collapse in 2D and 3D settings, respectively \cite%
{Berge,Fibich,Mardonov15}. Several possibilities for the stabilization of
the multidimensional solitons were theoretically developed, the most
ubiquitous one being the use of lattice potentials \cite%
{review1,review2,review3}. Recently, it was demonstrated that both 2D \cite%
{Sakaguchi14,Kart,Luca,XZZ,Sherman,SKA,Raymond,CNSNS,BLi,2018} and 3D \cite%
{HP} two-component solitons in the free space with the cubic self-focusing
can be stabilized with the help of spin-orbit coupling (SOC) in BEC (see a
summary in a recent brief review \cite{EPL}). The SOC effect is represented,
in experimentally implemented models of bosonic \cite%
{Bose-SOC1,Bose-SOC2,Bose-SOC3,erson,Bose-SOC4,Li2017} and fermionic \cite%
{Fermi-SOC,Fermi-SOC2} gases, by linear terms coupling the two components
through first spatial derivatives, see also reviews \cite%
{Spielman,Bose-SOC5,Gedeminas,Bose-SOC-review}. It is essential to stress
that, while most experimental works for SOC were carried out in effectively
1D settings, the realization of the SOC in 2D geometry was demonstrated too,
for bosons \cite{bosons-SOC-2D} and fermions \cite{fermions-SOC-2D} alike.
In 2D models, the SOC terms lift the specific conformal invariance of the
GPEs with cubic self-attraction, which is responsible for the onset of the
critical collapse. As a result, in this case the addition of the SOC to the
GPE system creates the otherwise missing ground state in the form of
semi-vortex (SV) solitons with one zero-vorticity (\textit{fundamental}) and
one vortical components, provided that the self-attraction is stronger than
the cross-attraction. In the opposite case, SVs are unstable, while the
ground state is represented by the mixed-mode (MM) state, which combines
fundamental and vortical terms in both components. Furthermore, the SVs and
MMs remain stable when, respectively, the cross- or self-interaction is
repulsive, the self-trapping being provided by the stronger self- (cross-)
attraction for SVs (MMs) \cite{Raymond}.

The aim of the present work is to obtain soliton solutions in the model of
an oblate (quasi-2D) Fermi superfluid, based on the known 2D
density-functional equations \cite%
{Yefsah2007,Antezza2007,Adhikari2008,Diaz2012,Yefsah2013,Jie2014,Diaz2015,Lombardi2017,Shang2018}%
, to which SOC terms are added. We demonstrate that the system produces
quiescent composite solitons in the free 2D space. Further, because the
Galilean invariance of the system is broken by the SOC terms, generating
moving solitons from the quiescent ones is a nontrivial issue too. We find
that moving solitons can be created, with velocities limited by a critical
value, which depends on the strength of the spin-orbit coupling as well as on soliton's norm. Collisions between moving
solitons are also addressed, by means of direct simulations.

The rest of the paper is organized as follows. The model is introduced in
Section \ref{SII}. Numerical results for quiescent and moving solitons are
presented in Sections \ref{SIII} and \ref{SIV}, respectively, the latter one
also including the consideration of collisions between moving solitons. The
paper is concluded by Section \ref{SVI}.

\begin{figure}[tbp]
\begin{centering}
\begin{tabular}{lll}
(a) & (b) &  \\
\includegraphics[width=0.24\textwidth]{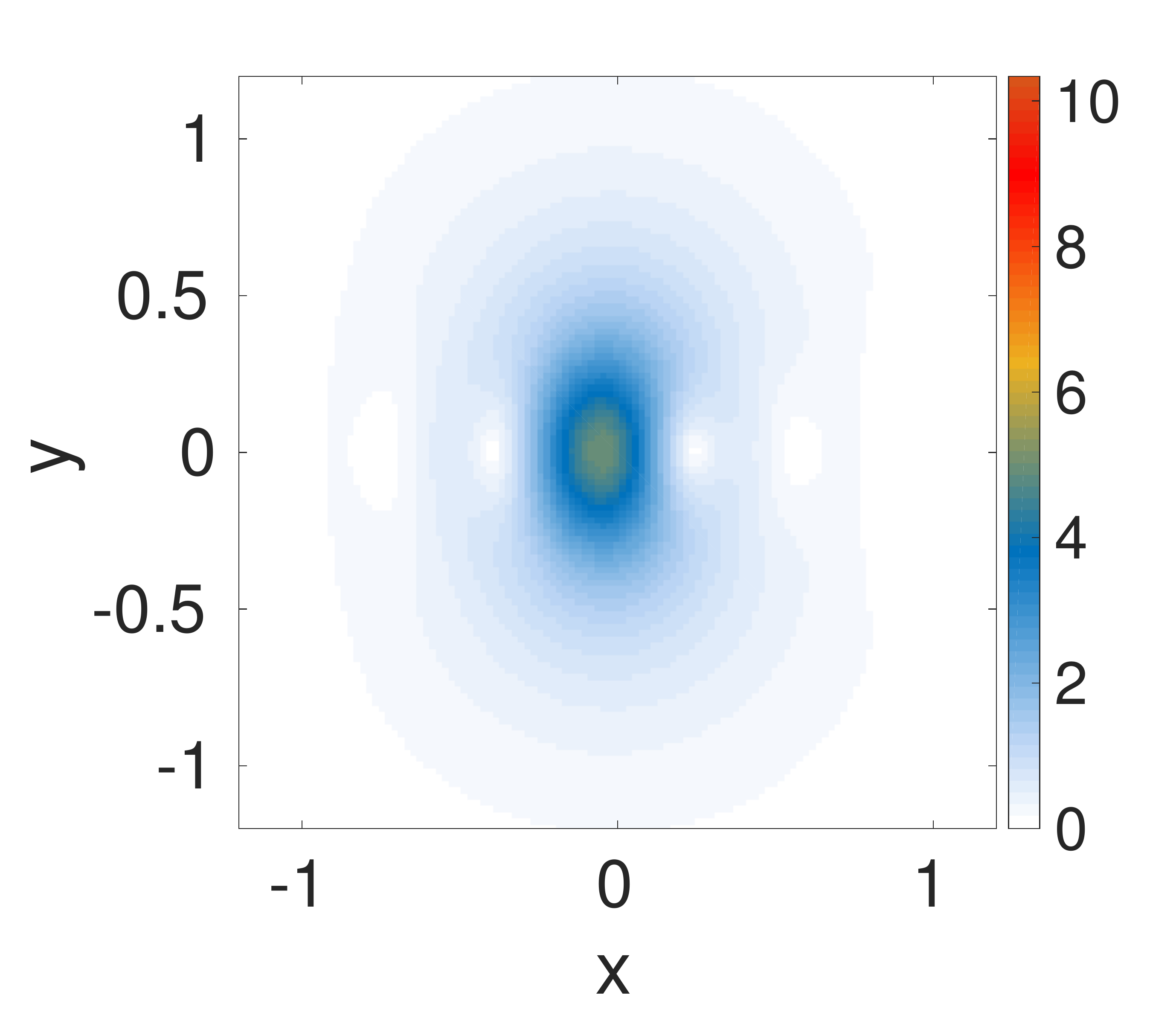} & %
\includegraphics[width=0.24\textwidth]{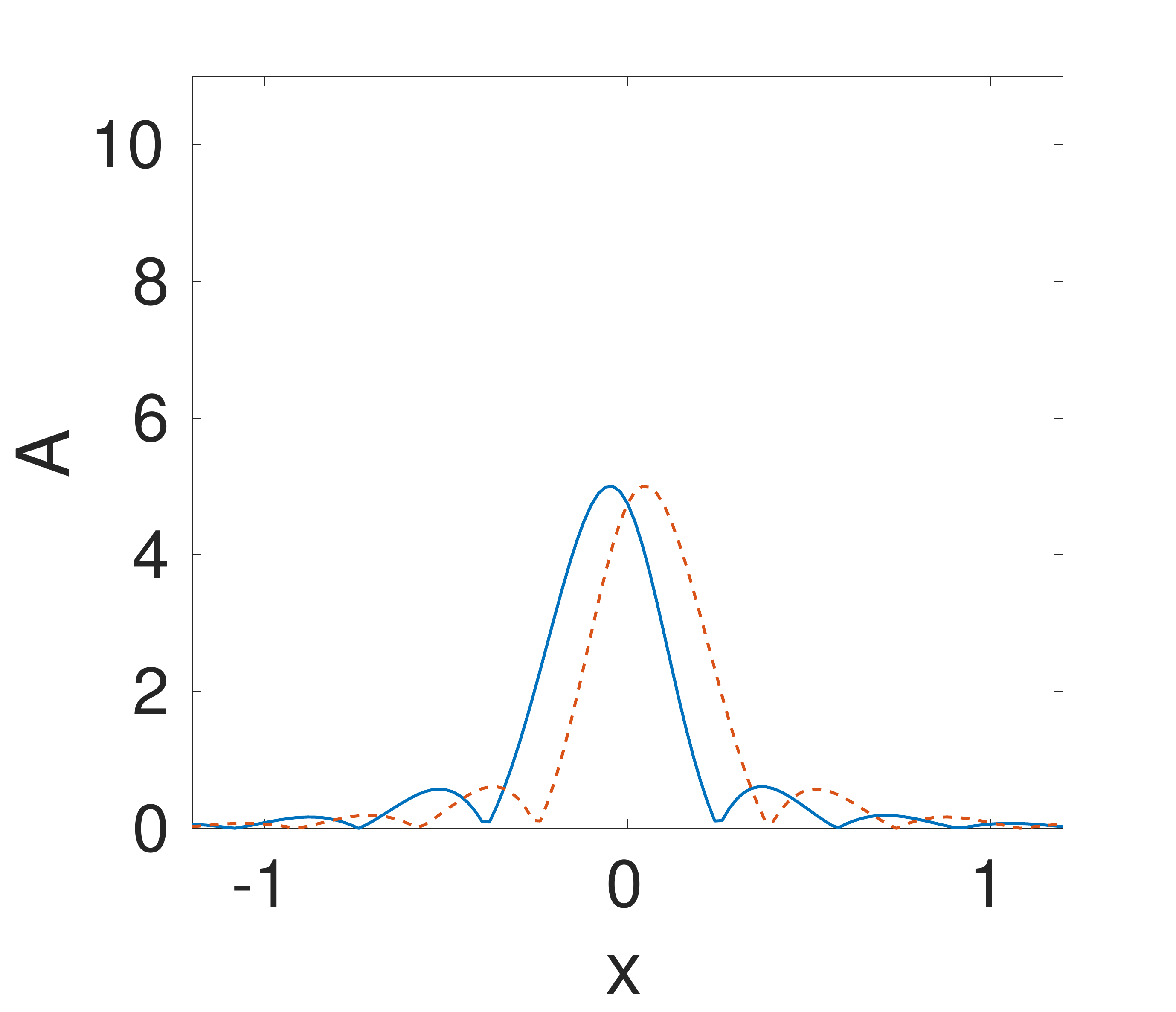} &  \\
&  &  \\
(c) & (d) &  \\
\includegraphics[width=0.24\textwidth]{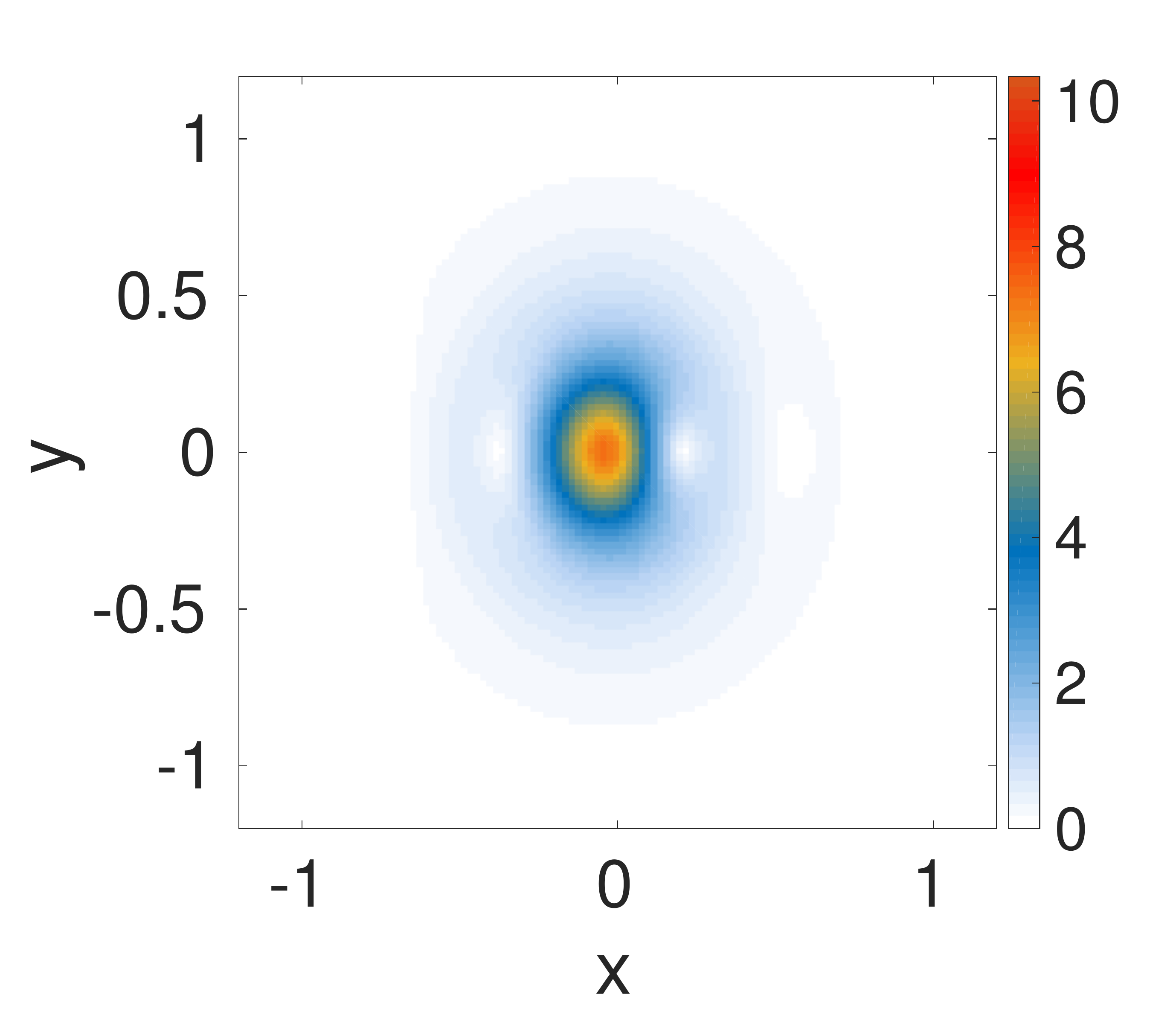} & %
\includegraphics[width=0.24\textwidth]{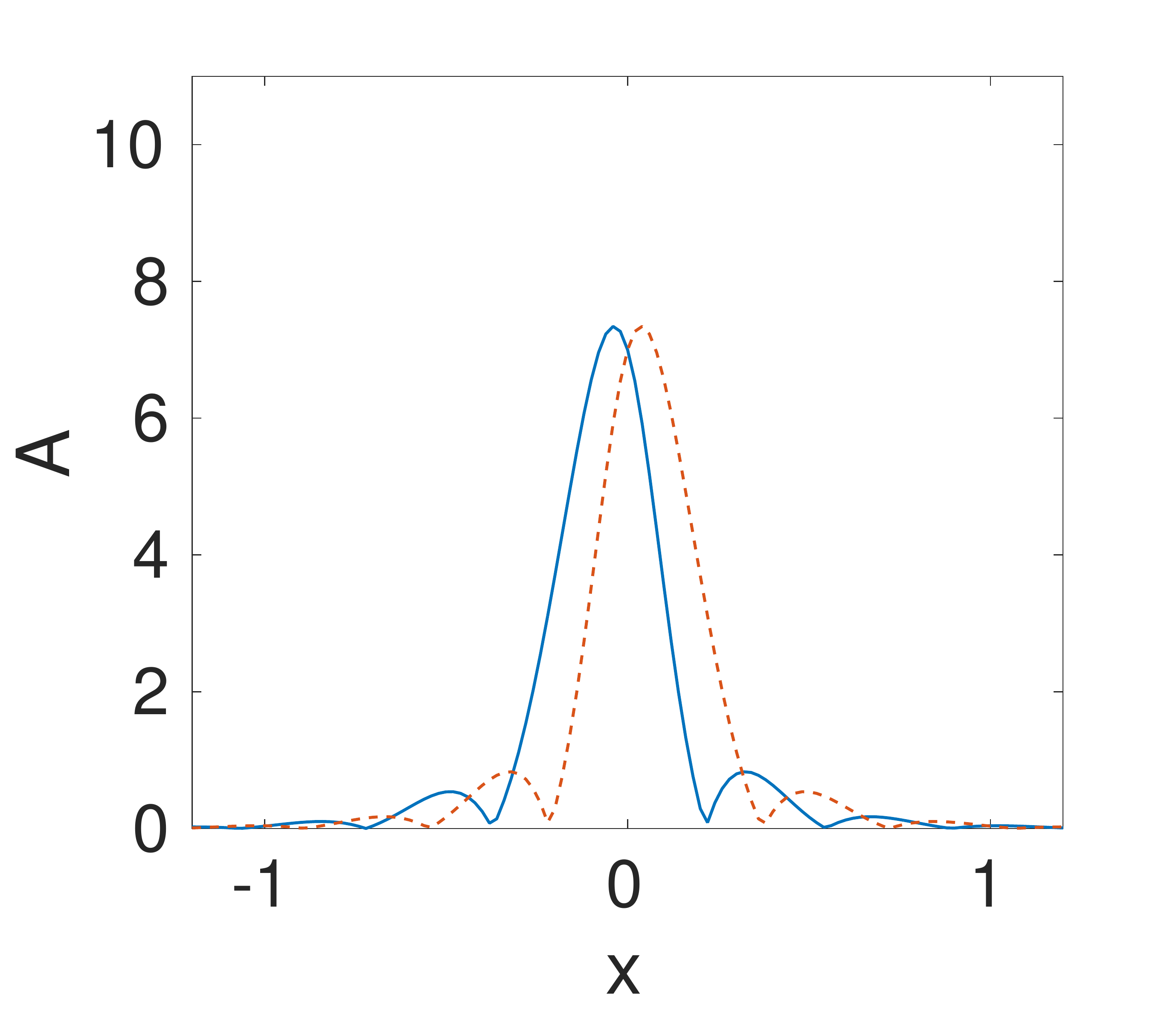} &  \\
&  &  \\
(e) & (f) &  \\
\includegraphics[width=0.24\textwidth]{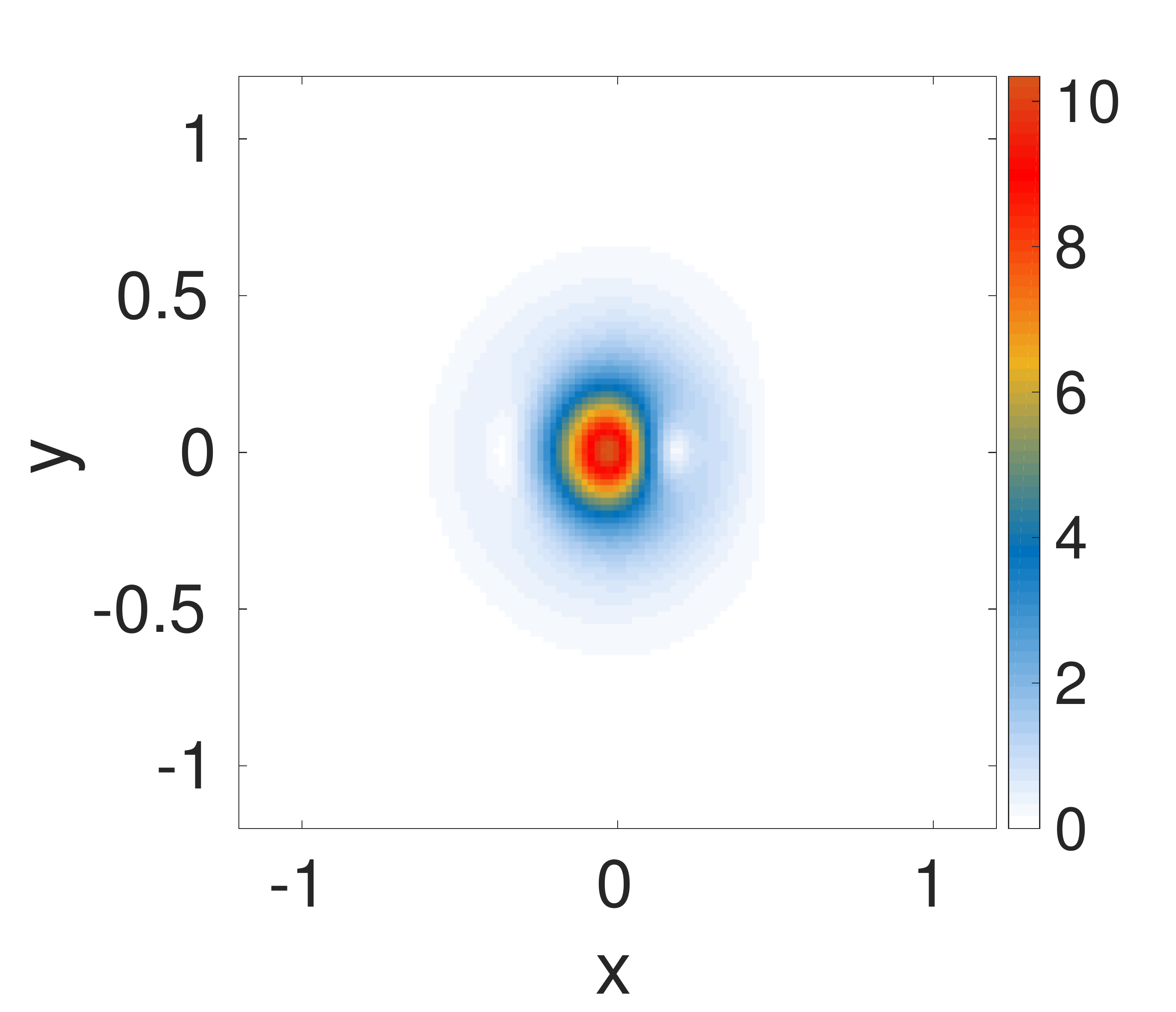} & %
\includegraphics[width=0.24\textwidth]{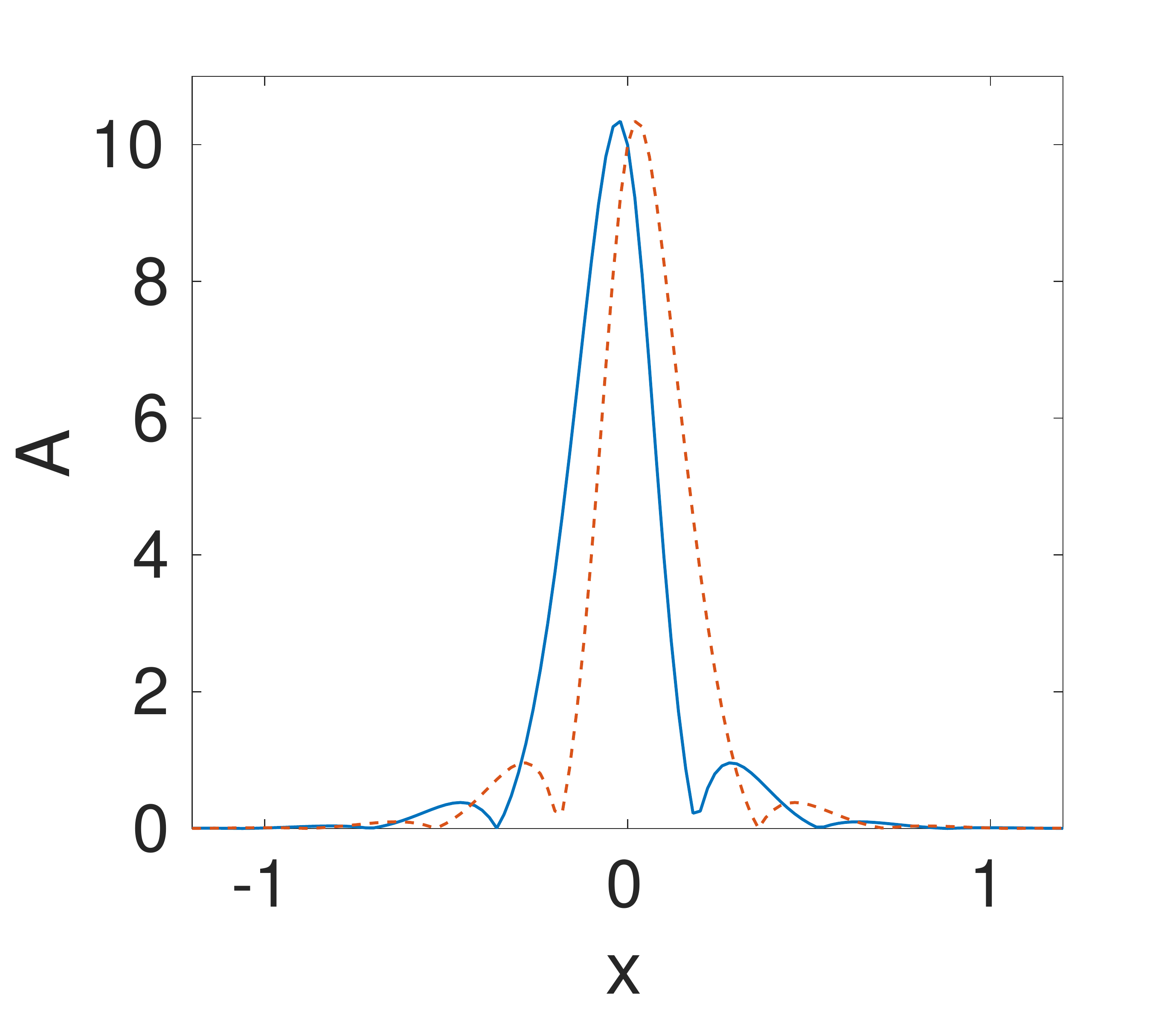} &  \\
&  &
\end{tabular}%
\caption{Left: contour plots of $|\Psi _{1}(\mathbf{r},t)|$ for stable
stationary MM solutions in the $\left( x,y\right) $ plane, as produced by
the full system based on Eqs. (\protect\ref{system}). Right: profiles of
both field components in the stationary solutions, $|\Psi _{1}(\mathbf{r}%
,t)| $ and $|\Psi _{2}(\mathbf{r},t)|$ (blue solid and red dotted lines,
respectively) in the cross-section $y=0$. Panels (a,b), (c,d) and (e,f)
correspond to total norms $N=8.0$, $9.5$, and $11.0$, respectively. Fixed
parameters are $\protect\gamma =1$, $\protect\varepsilon =0$ and $\protect%
\lambda =9$.}
\label{FIG1}
\end{centering}
\end{figure}

\section{The model}

\label{SII}

The system of 2D density-functional equations for the spinor wave function, $%
\Psi (\mathbf{r},t)=\left( \Psi _{1}(\mathbf{r},t),\Psi _{2}(\mathbf{r}%
,t)\right) $, of the binary Fermi gas, which include contact attraction
between the two components (spin-up and down atomic states) with strength $%
\gamma >0$, the effective self-repulsive Pauli nonlinearity, SOC terms of
the Rashba type \cite{Rashba,R&Sh} with strength $\lambda $, and Zeeman
splitting with strength $\varepsilon $ are written in the scaled form as 
\begin{equation}
i\frac{{\partial {\Psi _{1}}}}{{\partial t}} =-\frac{1}{2}{\nabla ^{2}}{\Psi _{1}}+{\left\vert {{\Psi _{1}}}\right\vert ^{4/3}}{\Psi _{1}}-\gamma {%
\left\vert {{\Psi _{2}}}\right\vert ^{2}}{\Psi _{1}}  +\lambda \left( {\frac{{\partial {\Psi _{2}}}}{{\partial x}}-i\frac{{\partial {\Psi _{2}}}}{{\partial y}}}\right) +\varepsilon {\Psi _{1}\,,} \notag
\end{equation}%
\begin{equation}
i\frac{{\partial {\Psi _{2}}}}{{\partial t}} =-\frac{1}{2}{\nabla ^{2}}{\Psi _{2}}+{\left\vert {{\Psi _{2}}}\right\vert ^{4/3}}{\Psi _{2}}-\gamma {\left\vert {{\Psi _{1}}}\right\vert ^{2}}{\Psi _{2}}  -\lambda \left( {\frac{{\partial {\Psi _{1}}}}{{\partial x}}+i\frac{{\partial {\Psi _{1}}}}{{\partial y}}}\right) -\varepsilon {\Psi _{2}.}
\label{system}
\end{equation}%
Starting from the full 3D system, one can derive the effectively 2D system by eliminating the third direction, in which the gas is assumed to be strongly confined by an external potential
(see, e.g., work \cite{AM06}). Along with the Hamiltonian and 2D momentum,
the system conserves the total norm, which is proportional to the number of
atoms in the degenerate gas,
\begin{equation}
N=\int {\int {\left( {{{\left\vert {{\Psi _{1}}\left( {\mathbf{r},t}\right) }%
\right\vert }^{2}}+{{\left\vert {{\Psi _{2}}\left( {\mathbf{r},t}\right) }%
\right\vert }^{2}}}\right) }}{d^{2}}r.  \label{N}
\end{equation}%
It is relevant to stress that the local interaction between the polarized
spin-up and down fermions may be controlled, as concerns its sign and
strength, by means of the Feshbach resonance \cite{Feshbach-Fermi,Feshbach}.

In the limit case or of large atomic density, $\gamma \left\vert \Psi
_{1,2}\right\vert ^{2/3}\gg 1$, the Pauli self-repulsion may be neglected,
reducing Eq. (\ref{system}) to%
\begin{gather}
i\frac{{\partial {\Psi _{1}}}}{{\partial t}}=-\frac{1}{2}{\nabla ^{2}}{\Psi
_{1}}-\gamma {\left\vert {{\Psi _{2}}}\right\vert ^{2}}{\Psi _{1}}   
+\lambda \left( {\frac{{\partial {\Psi _{2}}}}{{\partial x}}-i\frac{{%
\partial {\Psi _{2}}}}{{\partial y}}}\right) +\varepsilon {\Psi _{1},}
\notag
\end{gather}%
\begin{gather}
i\frac{{\partial {\Psi _{2}}}}{{\partial t}}=-\frac{1}{2}{\nabla ^{2}}{\Psi
_{2}}-\gamma {\left\vert {{\Psi _{1}}}\right\vert ^{2}}{\Psi _{2}}  -\lambda \left( {\frac{{\partial {\Psi _{1}}}}{{\partial x}}+i\frac{{%
\partial {\Psi _{1}}}}{{\partial y}}}\right) -\varepsilon {\Psi _{2}.}
\label{limitc}
\end{gather}%
The same system (\ref{limitc})\ applies to a binary bosonic condensate
dominated by the inter-species attraction, which may also be enhanced by
means of the Feshbach resonance \cite{Raymond}.

Using the remaining scaling invariance of Eqs. (\ref{system}) and (\ref%
{limitc}), we fix $\gamma =1$ (the case of $\gamma =0$ is irrelevant here,
as in that case the system cannot produce bright solitons), while the SOC
and Zeeman strengths $\lambda $ and $\varepsilon $ remain irreducible
parameters in the full system (\ref{system}). In the reduced system (\ref%
{limitc}), $\lambda $\ it may be scaled to any fixed value, the only
remaining free parameter being $\varepsilon $.

The spectrum of small-amplitude plane waves, generated by the linearization
of Eqs. (\ref{system}) or (\ref{limitc}) for $\Psi _{1,2}\sim \exp \left( i%
\mathbf{k}\cdot \mathbf{r}-i\mu t\right) $, where $\mathbf{k}$ is the wave
vector and $\mu $ a real chemical potential, gives rise to two branches \cite%
{NJP}:
\begin{equation}
\mu =\frac{k^{2}}{2}\pm \sqrt{\lambda ^{2}k^{2}+\varepsilon ^{2}}.
\label{eps}
\end{equation}%
Solitons may exist at values of $\mu $ which are not covered by Eq. (\ref%
{eps}) with real $k^{2}\geq 0$, i.e., at%
\begin{gather}
\mu <-\frac{1}{2}\left( \lambda ^{2}+\frac{\varepsilon ^{2}}{\lambda ^{2}}%
\right) ,~\mathrm{if}~~\lambda ^{2}>|\varepsilon |,  \notag \\
\mu <-|\varepsilon |,~\mathrm{if}~~\lambda ^{2}<|\varepsilon |.  \label{mu}
\end{gather}

Stationary states with chemical potential $\mu $ are sought for as
\begin{equation}
\Psi _{1,2}=e^{-i\mu t}U_{1,2}\left( x,y\right) ,  \label{U}
\end{equation}%
where complex amplitude functions $U_{1,2}$ satisfy equations
\begin{gather*}
\mu U_{1}=-\frac{1}{2}{\nabla ^{2}U_{1}}+{\left\vert {{U_{1}}}\right\vert
^{4/3}U_{1}}-\gamma {\left\vert {{U_{2}}}\right\vert ^{2}U_{1}} 
+\lambda \left( {\frac{{\partial U{_{2}}}}{{\partial x}}-i\frac{{\partial U{%
_{2}}}}{{\partial y}}}\right) +\varepsilon {U_{1},}
\end{gather*}%
\begin{gather}
\mu U_{2}=-\frac{1}{2}{\nabla ^{2}U_{2}}+{\left\vert {{U_{2}}}\right\vert
^{4/3}U_{2}}-\gamma {\left\vert {{U_{1}}}\right\vert ^{2}U_{2}} 
-\lambda \left( {\frac{{\partial U{_{1}}}}{{\partial x}}+i\frac{{\partial U{%
_{1}}}}{{\partial y}}}\right) -\varepsilon {U_{2}.}  \label{UU}
\end{gather}%
We have produced stationary states, solving Eqs. (\ref{system}) and (\ref%
{limitc}) by means of the well-known imaginary-time method \cite{IT1,IT2},
and stability of the stationary states was then tested by means of real-time
simulations of their perturbed evolution. This was done by means of the
fourth-order Runge-Kutta method for marching in time with step $\Delta
t=10^{-4}$, handling the spatial derivatives by means of the centered
second-order finite-difference method with $\Delta x=\Delta y=0.025$, in
terms of the scaled variables. The numerical solutions were obtained in a 2D
domain, defined as $|x|,|y|<4$, with zero boundary conditions, the size of
the solitons always being much smaller than the domain width. {Reliability
of the numerical results was checked by reproducing them with smaller values
of $\Delta t$ and $\Delta x$, $\Delta y$, as well as with different sizes of
the integration domain.}

\section{Stationary mixed-mode solitons}

\label{SIII}

\subsection{The full model}

\begin{figure}[tbp]
\begin{centering}
\begin{tabular}{lll}
(a) & (b) &  \\
\includegraphics[width=0.24\textwidth]{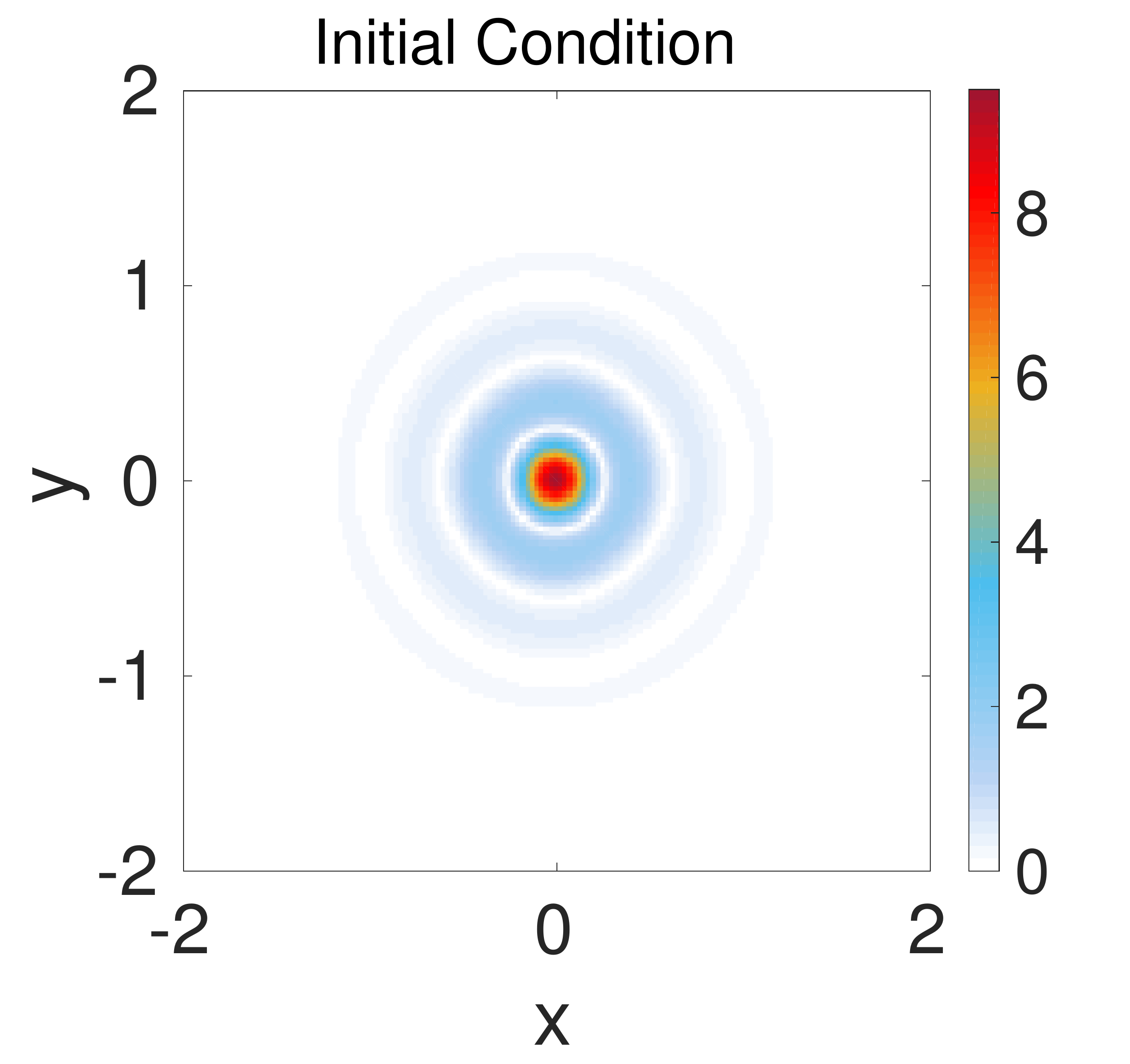} & %
\includegraphics[width=0.24\textwidth]{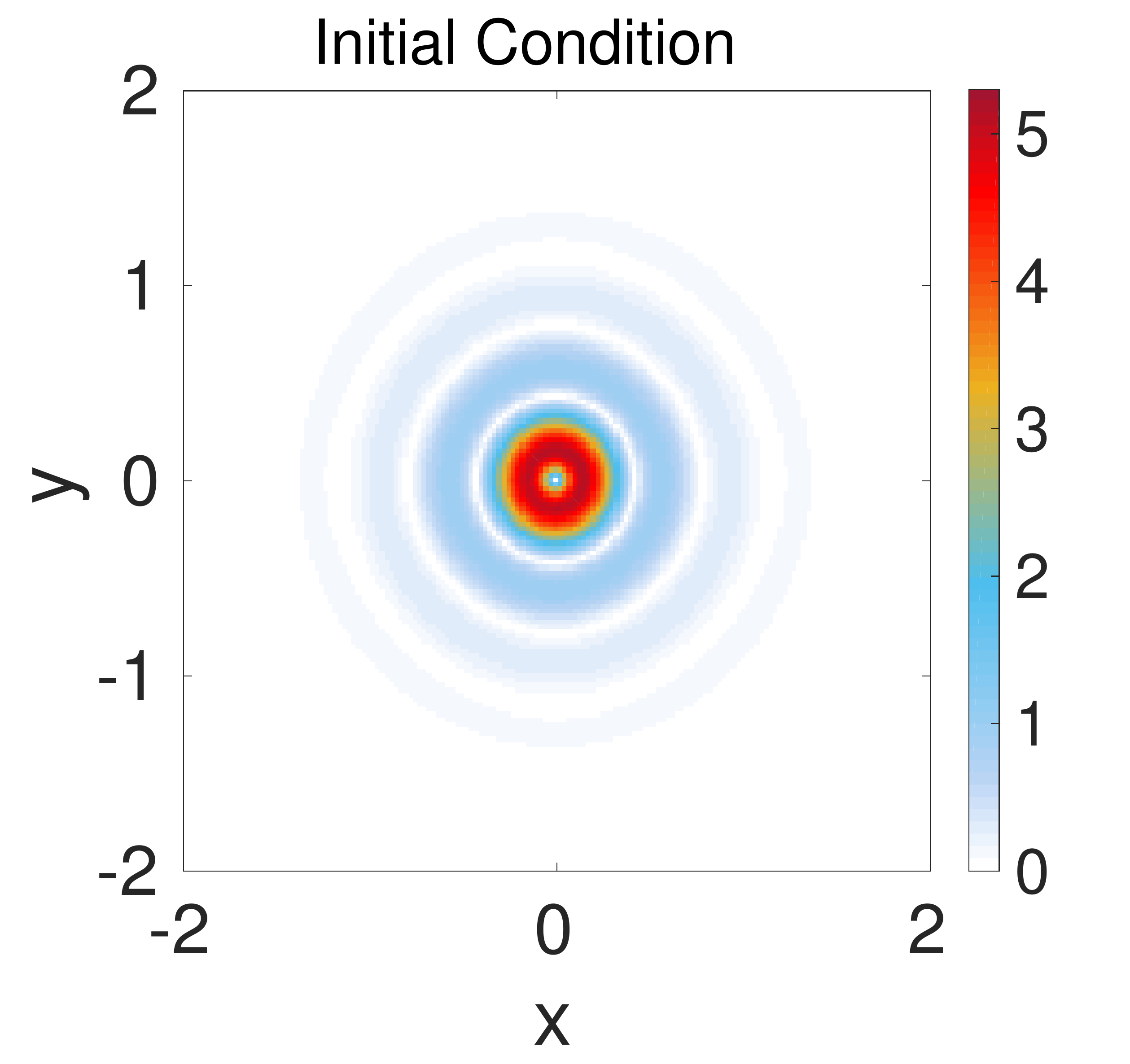} &  \\
&  &  \\
(c) & (d) &  \\
\includegraphics[width=0.24\textwidth]{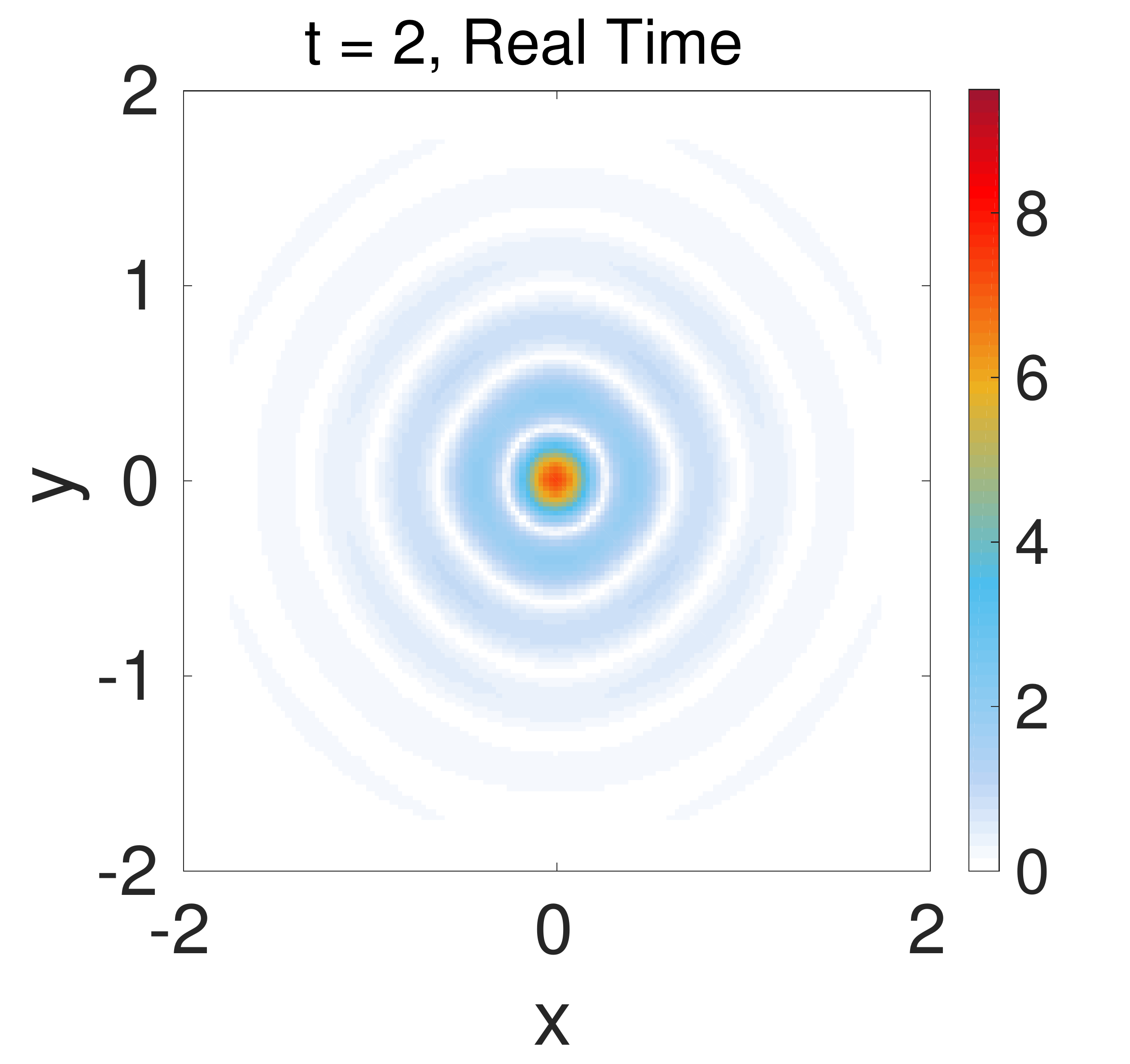} & %
\includegraphics[width=0.24\textwidth]{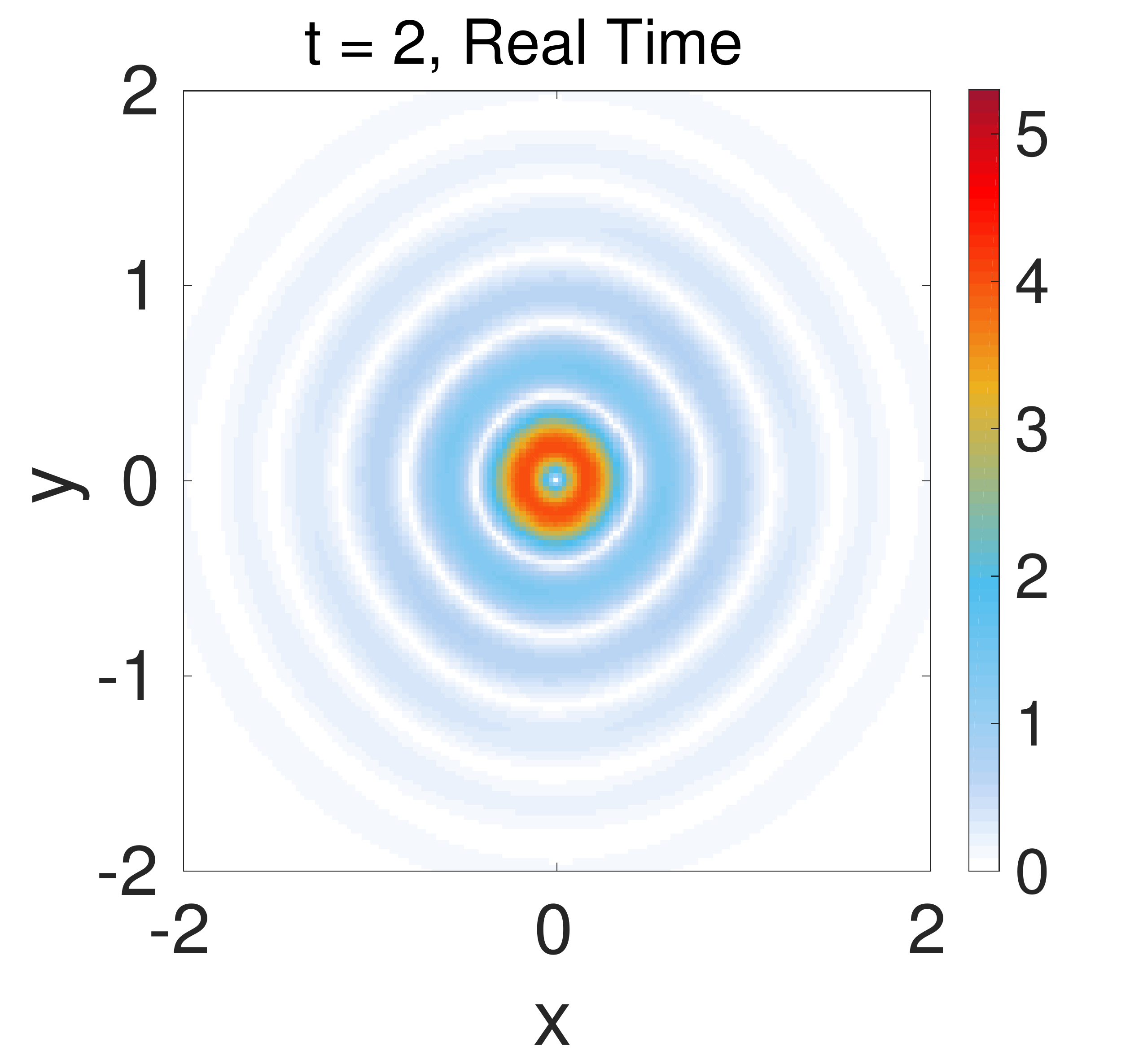} &  \\
&  &  \\
(e) & (f) &  \\
\includegraphics[width=0.24\textwidth]{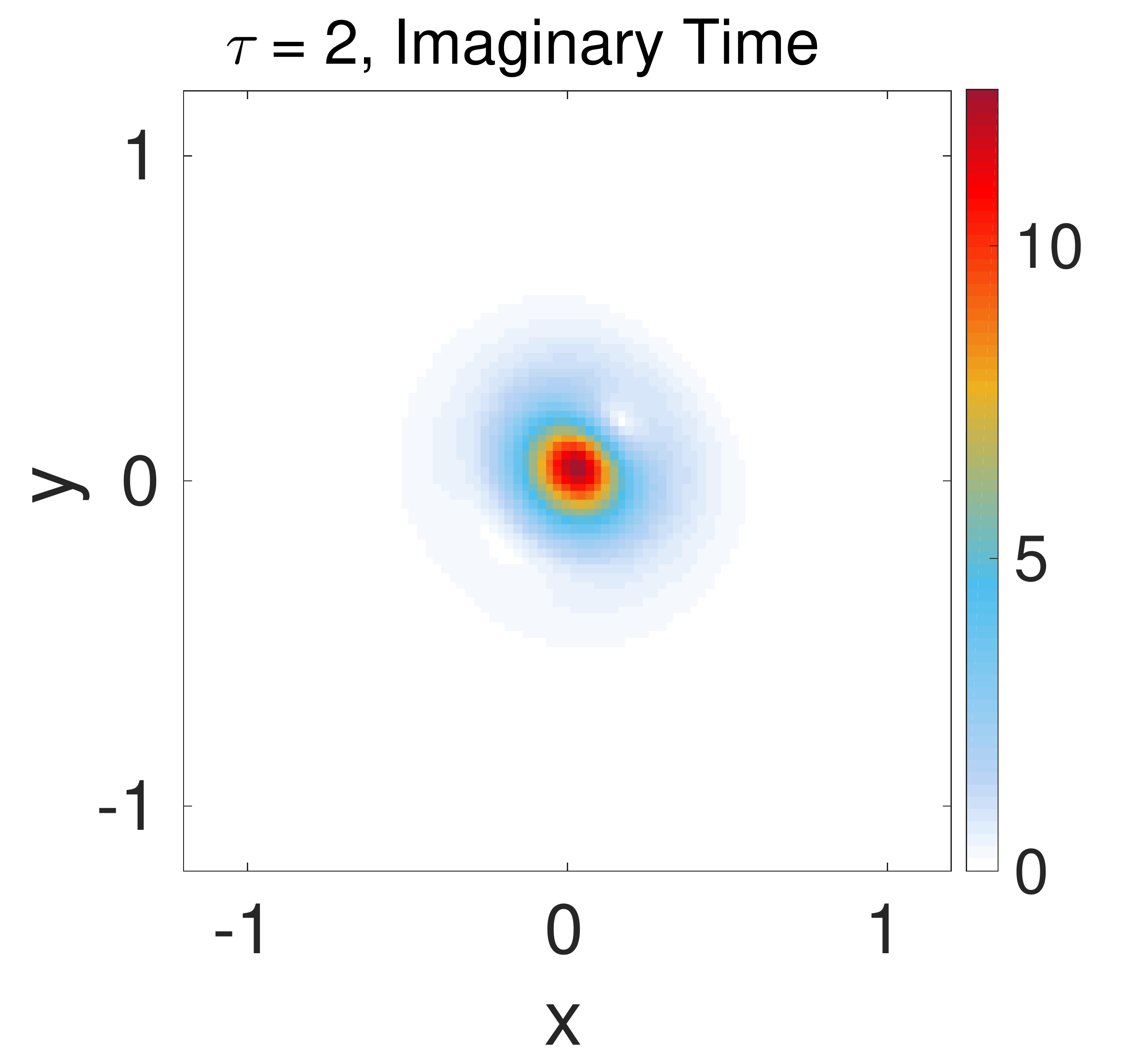} & %
\includegraphics[width=0.24\textwidth]{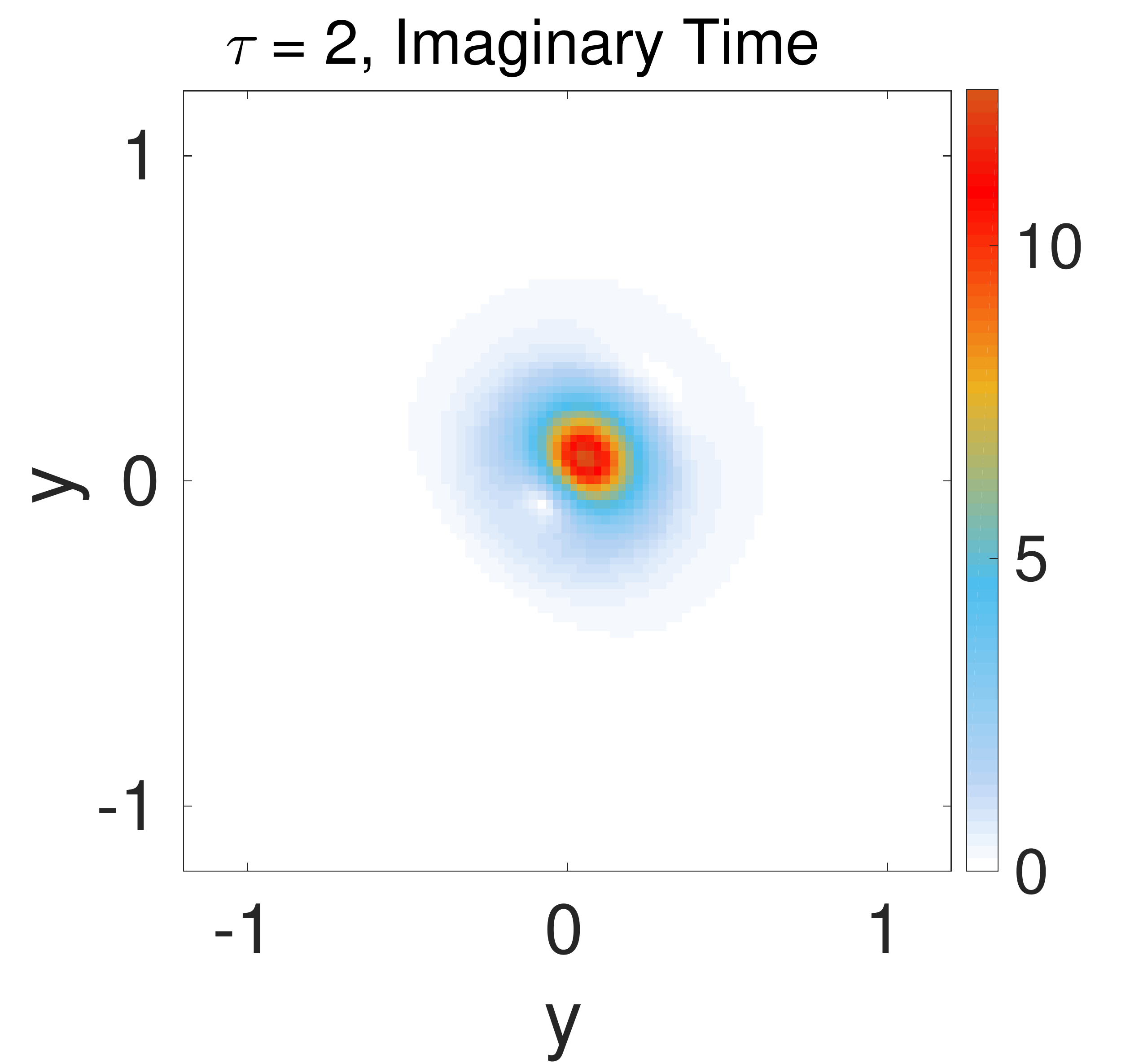} &  \\
&  &
\end{tabular}%
\caption{Contour plots of $|\Psi _{1}(\mathbf{r},t)|$ and $|\Psi _{2}(%
\mathbf{r},t)|$ (left and right panels, respectively) of an unstable SV. $(%
\mathrm{a,b})$ The SV state produced by the evolution in imaginary time, $t=i%
\protect\tau $, from input (\protect\ref{SV-input}), at $\protect\tau =27.5$%
. $(\mathrm{c,d})$ The result of the unstable evolution in real time from $%
t=0$ to $t=2$. $(\mathrm{e,f})$ The result of the additional evolution of
the SV from panels $\left( \mathrm{a,b}\right) $ in the imaginary time, $%
\Delta t=i\Delta \protect\tau $, at $\Delta \protect\tau =2$, which
demonstrates spontaneous rearrangement of the SV into a stable MM. Fixed
parameters are $N=11.5$, $\protect\gamma =1$, $\protect\varepsilon =0$ and $%
\protect\lambda =9$.}
\label{FIGNEW}
\end{centering}
\end{figure}

The analysis is initiated by dropping the Zeeman splitting, i.e., setting $%
\varepsilon =0$, but keeping the Pauli-repulsion terms in Eqs. (\ref{system}%
). In this system, self-trapped states of the MM type were generated by
imaginary-time simulations, initiated with the input combining vorticities $%
(0,-1)$ and $(0,+1)$ in the two components, cf. Ref. \cite{Sakaguchi14}:%
\begin{equation}
\Psi _{1}^{(0)}(\mathbf{r})=A_{1}\exp (-\alpha _{1}r^{2})-A_{2}r\exp
(-i\theta -\alpha _{2}r^{2}),  \label{input1} \notag
\end{equation}%
\begin{equation}
\Psi _{2}^{(0)}(\mathbf{r})=A_{1}\exp (-\alpha _{1}r^{2})+A_{2}r\exp
(i\theta -\alpha _{2}r^{2}).  \label{input2}
\end{equation}%
Subsequently, real-time simulations reveal the stability of the stationary
MMs in their \emph{full existence domain}, in both the full and reduced
models, based on Eqs. (\ref{system}) and (\ref{limitc}), respectively.
Figure \ref{FIG1} shows typical contour plots and cross-section profiles of
stationary solitons of the MM type, obtained for three different values of
norm (\ref{N}). Naturally, as $N$ increases, making the cross-attraction
stronger, the solitons' amplitude increases too.

In addition, 2D solitons of the SV type were produced too, with an input
carrying vorticity $0$ in one component and $1$ in the other, as suggested
by work \cite{Sakaguchi14}. In this case, Eqs. (\ref{input1}) and (\ref%
{input2}) are reduced to
\begin{equation}
\Psi _{1}^{(0)}(\mathbf{r})=A_{1}\exp (-\alpha _{1}r^{2}), \notag
\end{equation}
\begin{equation}
\Psi _{2}^{(0)}(%
\mathbf{r})=A_{2}r\exp (i\theta -\alpha _{2}r^{2}).  \label{SV-input}
\end{equation}
However in contrast to the MMs, the so built SVs are \emph{completely
unstable}, which resembles the result for the bosonic binary condensate,
obtained in work \cite{Raymond}, in the case of the repulsive
self-interaction and attractive cross-interaction, combined with SOC.

The instability of the SV is illustrated in Fig. \ref{FIGNEW}, which
displays the evolution of its two components. In particular, panels (a,b)
show the result of the evolution in the imaginary time, $t=i\tau $, of input
(\ref{SV-input}), at $\tau =27.5$. The SV\ shape is maintained by the
evolution up to this value of $\tau $. However, if this SV is used as the
initial state for the real-time simulations, panels (c,d) show that the
result is instability, which causes decay of the SV. Further, if the
imaginary-time simulation is continued from $\tau =27.5$, which corresponds
to panels (a,b), by adding $\Delta \tau =2$, the ostensibly stable SV
spontaneously rearranges into a truly stable MM, see panels (e,f) (the value
of the Hamiltonian corresponding to the emerging MM is $H_{\mathrm{MM}%
}=-750.1$, which is smaller than the SV's original value, $H_{\mathrm{SV}%
}=-475.3$, which confirms that the MM is stable, while the SV is not). The
stability of this MM is confirmed by real-time simulations (not shown here
in detail).

\begin{figure}[tbp]
\begin{centering}
\begin{tabular}{lll}
(a) & (b) &  \\
\includegraphics[width=0.24\textwidth]{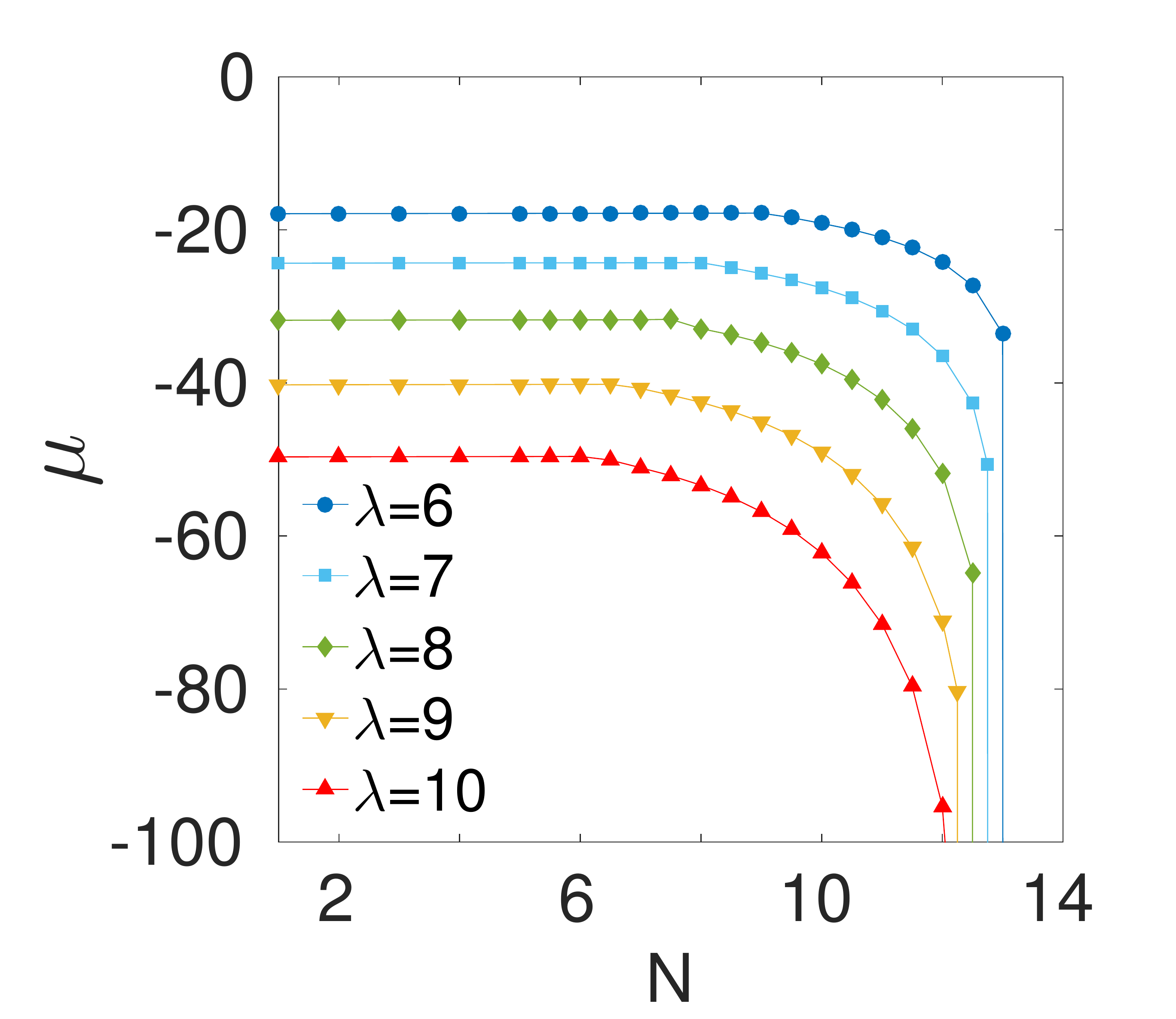} & %
\includegraphics[width=0.24\textwidth]{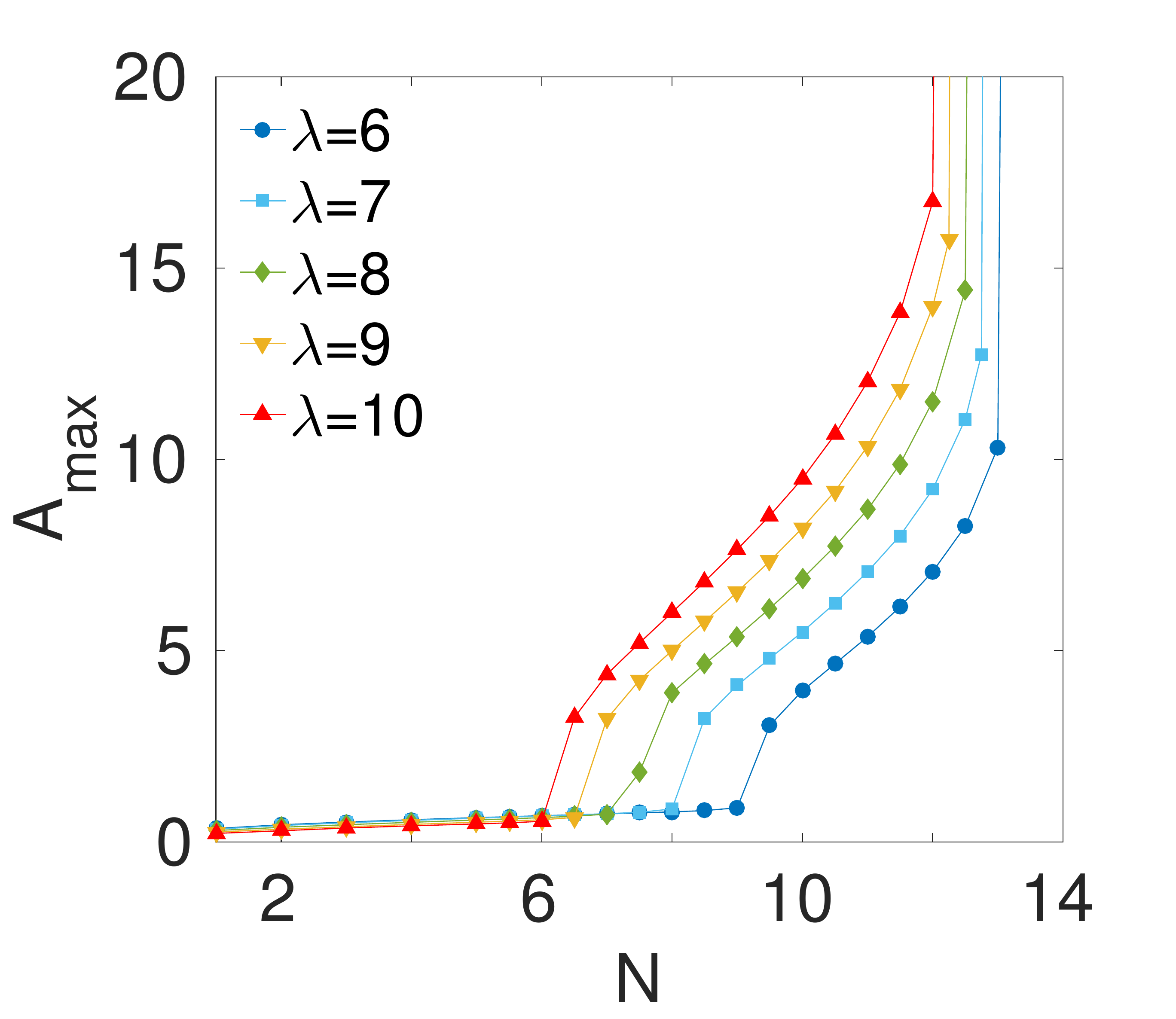} &  \\
&  &
\end{tabular}%
\caption{Chemical potential, $\protect\mu $, and the amplitude, $A_{\max }$
of stable mixed modes in the full system (\protect\ref{system}) versus the
norm, $N$, for different fixed values of the SOC strength, $\protect\lambda $%
. The drop of $A_{\max }$ to very small values at $N=N_{\min }$ in (b)
(e.g., $N_{\min }\approx 6$ at $\protect\lambda =10$) implies that the MM
soliton suffers delocalization into a state which tends to occupy the entire
integration domain). Other fixed parameters are the same as in Fig. \protect
\ref{FIG1}: $\protect\gamma =1$, $\protect\varepsilon =0$.}
\label{FIG2}
\end{centering}
\end{figure}

Figure \ref{FIG2} displays the dependence of the MM's chemical potential, $%
\mu $, and amplitude, $A_{\max }$ [the largest value of $\left\vert
U_{1,2}(x,y)\right\vert $, in Eq. (\ref{U})], on norm $N$ and SOC strength $%
\lambda $. The stability of the MMs complies with the fact that all the $\mu
(N)$ curves satisfy the Vakhitov-Kolokolov criterion, which is a well-known
necessary condition for the existence of all self-trapped states created by
attractive interactions \cite{VaKo,Berge,Fibich}. Further, Fig. \ref{FIG2}%
(b) clearly shows that, for given SOC strength $\lambda $, the MM exists at $%
N>N_{\min }$ (where $N_{\min }$ is the value of $N$ at which the amplitude
drops to very small values in Fig. Fig. \ref{FIG2}(b)), and vice versa: for
given $N$, it exists at $\lambda $ exceeding a certain finite minimum value,
$\lambda _{\min }$. At $N<N_{\min }$, solitons cannot self-trap in the full
system (\ref{system}), as in that case the Pauli self-repulsion in each
component dominates over the inter-component attraction.

\begin{figure}[tbp]
\begin{centering}
\includegraphics[width=0.4\textwidth]{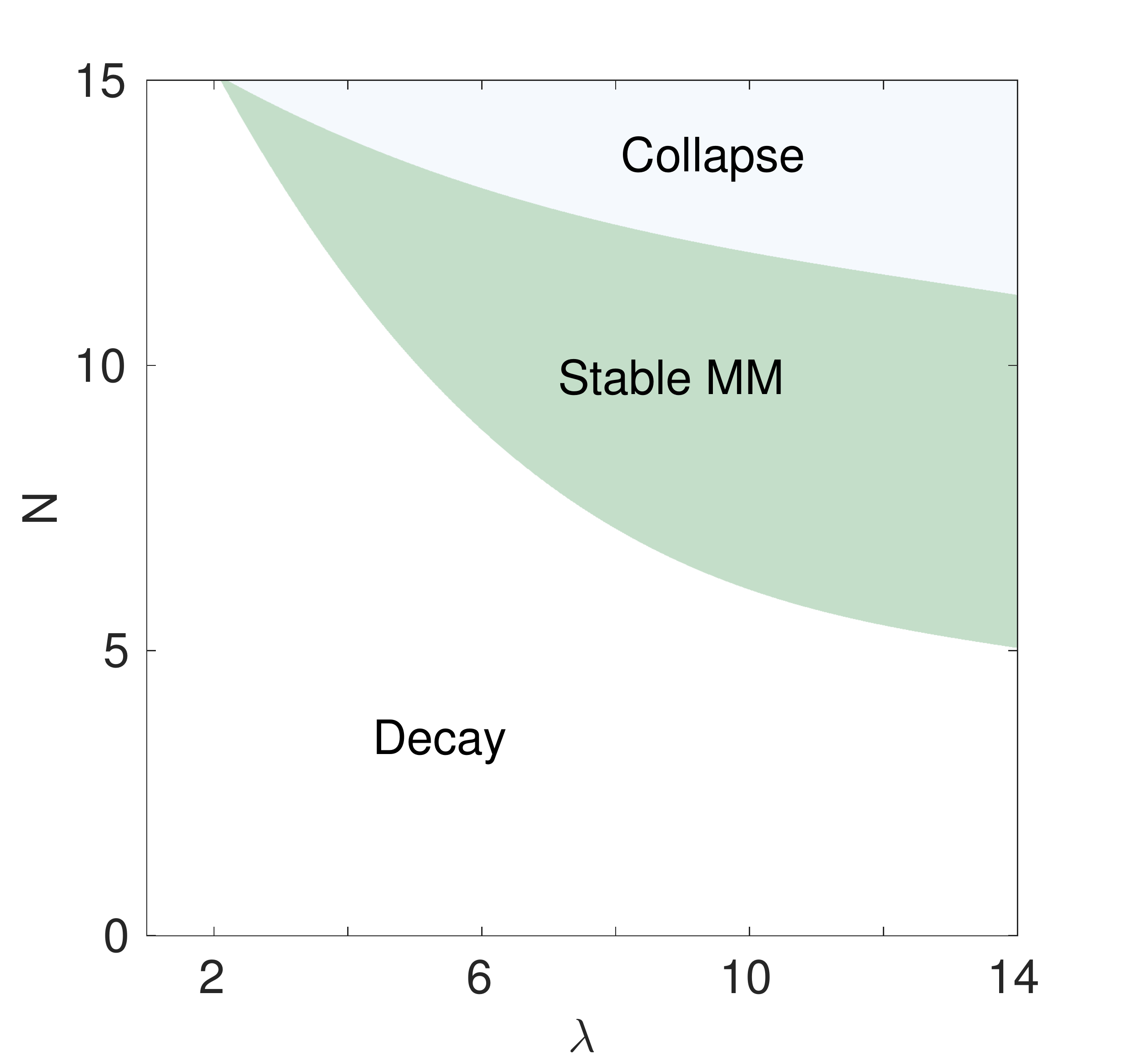}
\caption{The stability area for MMs in the plane of the SOC strength, $%
\protect\lambda $, and total norm, $N$. In the \textquotedblleft collapse"
area, the cubic cross-attraction is strong enough to drive the system
towards the collapse. MM-shaped inputs decay below the stability area, where
the attractive interaction is insufficient to secure the self-trapping.}
\label{phase}
\end{centering}
\end{figure}

To summarize these findings, Fig. \ref{phase} shows the MM stability region
in the $\left( {\lambda ,N}\right) $ plane. In particular, the nonexistence
of MMs at too small and too large values of $N$, observed in the latter
figure, is consistent with the same features demonstrated by Fig. \ref{FIG2}%
(b).

\begin{figure}[tbp]
\centering
\resizebox{0.4\textwidth}{!}{\includegraphics{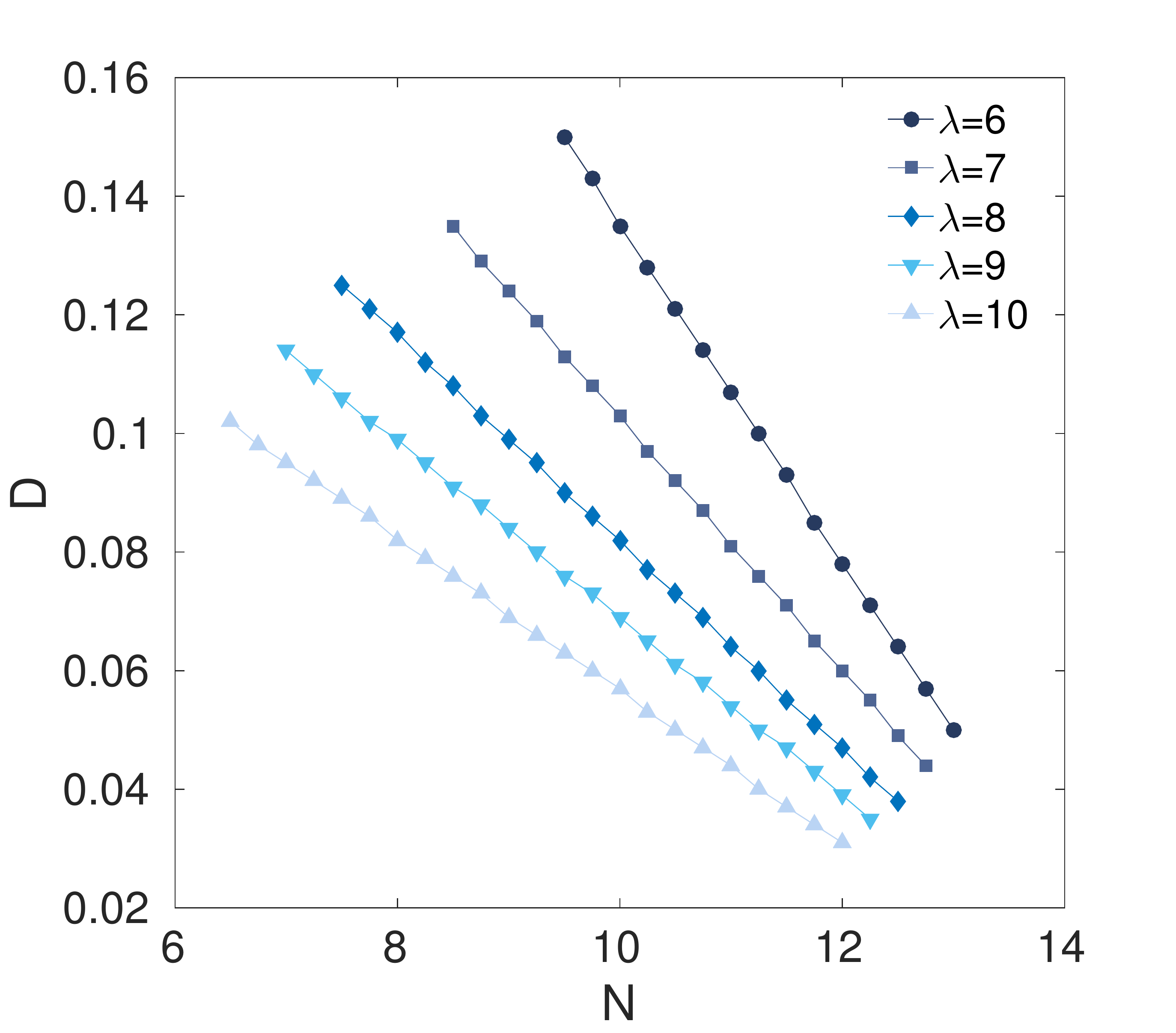}}
\caption{Separation $D$ (in the $x$ direction) between positions of density
maxima in the two components of the mixed modes, as a function $N$ for
different fixed values of $\protect\lambda $. Other parameters are the same
as in Fig. \protect\ref{FIG1}.}
\label{FIG3}
\end{figure}

Similar to the situation analyzed in the binary BEC \cite{Sakaguchi14}, the
MM structure is characterized by spatial separation, $D$, in the $x$
direction, between density peaks of the two components. $D$ is plotted, as a
function of $N$, for different fixed values of $\lambda $ in Fig. \ref{FIG3}%
. Naturally, the increase of $N$ strengthens the attraction between the
components, thus leading to decrease of $D$.

\begin{figure}[tbp]
\centering
\begin{tabular}{lll}
(a) & (b) &  \\
\includegraphics[width=0.24\textwidth]{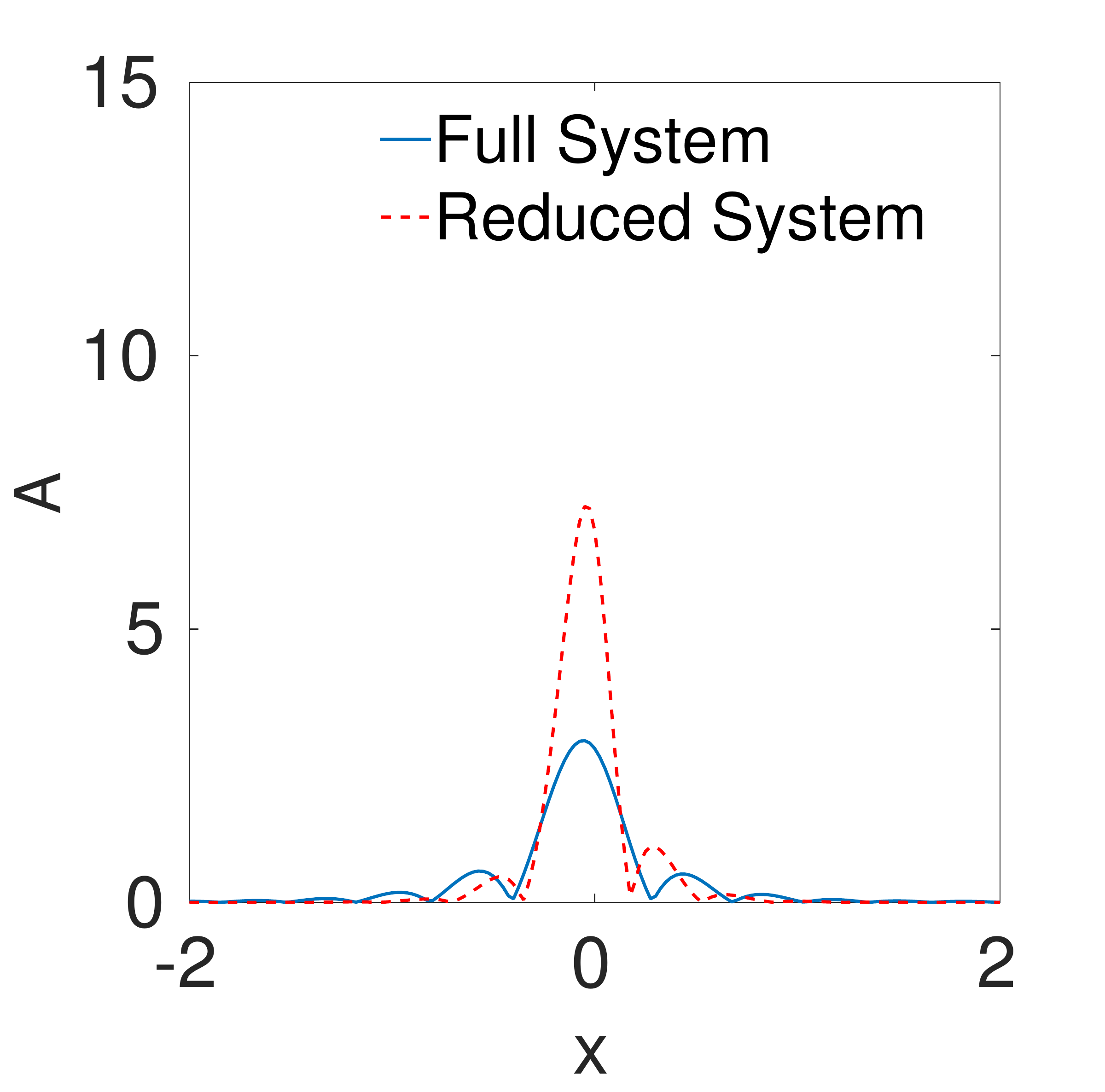} & %
\includegraphics[width=0.24\textwidth]{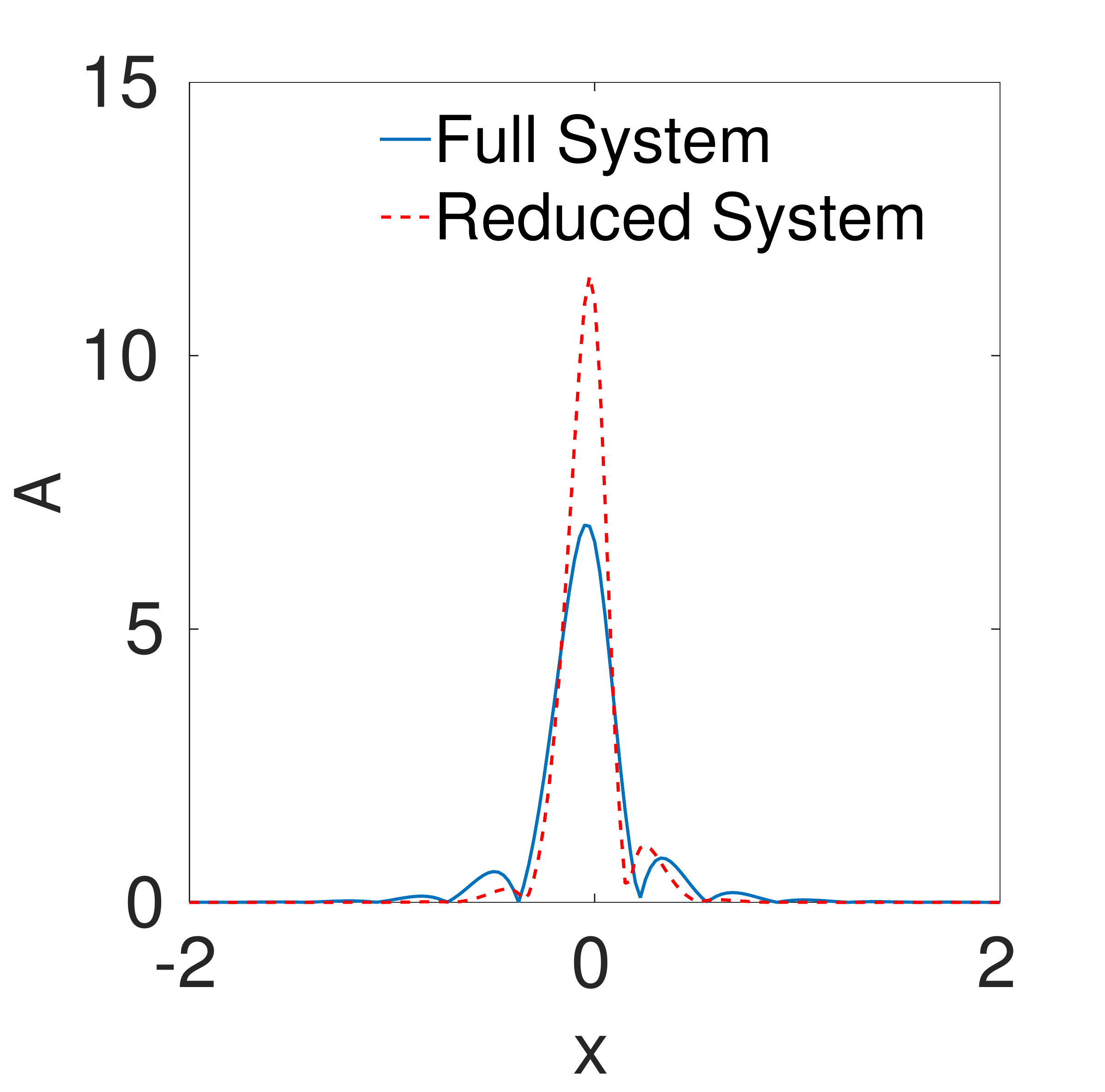} &  \\
&  &
\end{tabular}%
\caption{Comparison of cross-section profiles of $|\Psi _{1}(x)|$ at $y=0$
between the MM solitons produced by the the full and reduced systems (solid
and dashed lines, respectively). The profiles are produced for $N=7.0$ (a)
and $N=9.25$ (b). The other parameters are the same as in Fig. \protect\ref%
{FIG1}.}
\label{FIG4}
\end{figure}

\subsection{The reduced system and one with the Zeeman splitting}

\begin{figure}[tbp]
\centering
\begin{tabular}{lll}
(a) & (b) &  \\
\includegraphics[width=0.24\textwidth]{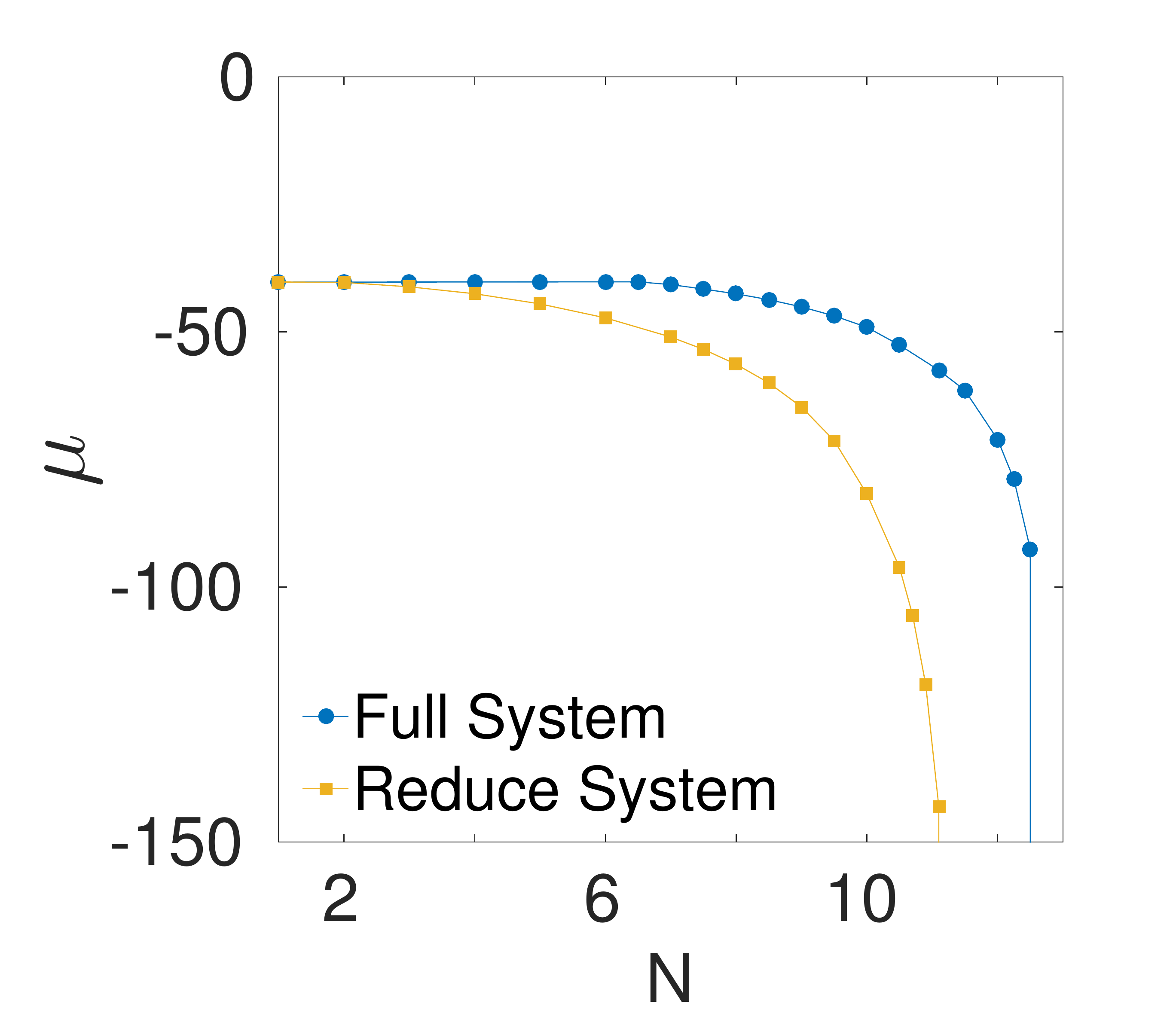} & %
\includegraphics[width=0.24\textwidth]{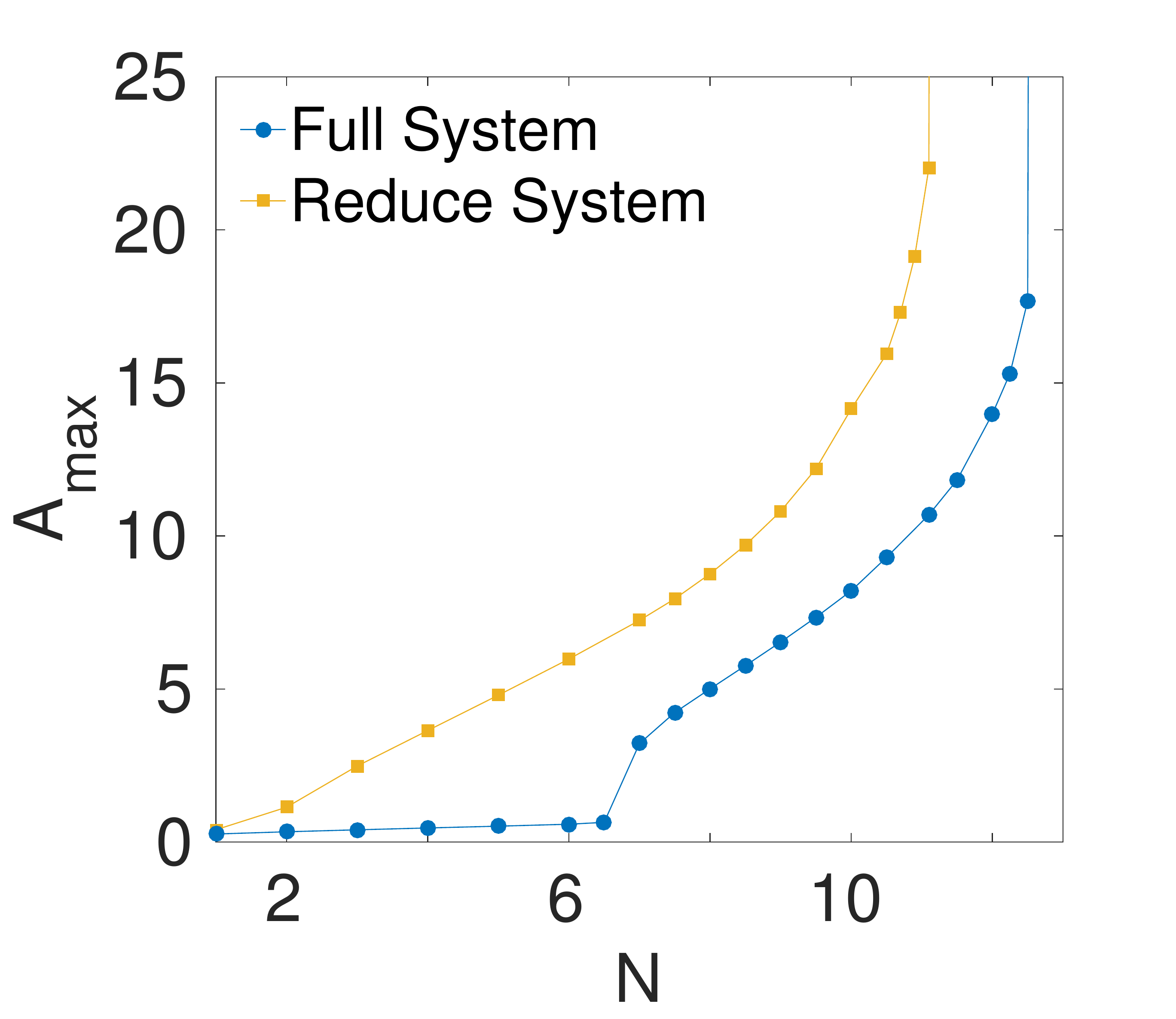} &  \\
&  &
\end{tabular}%
\caption{Comparison of the curves for the chemical potential, $\protect\mu %
(N)$, and amplitude, $A_{\max }(N)$, obtained in the full and reduced
systems (rhombuses and triangles, respectively). Other fixed parameters are
the same as in Fig. \protect\ref{FIG1}.}
\label{FIG5}
\end{figure}

Comparison of cross sections of the MM solitons found from Eq. (\ref{limitc}%
), which does not include the Pauli-repulsion terms (as mentioned above, it
applies to the binary BEC as well), with their counterparts produced by the
full system (\ref{system}) is shown in Fig. \ref{FIG4}. Due to the absence
of the repulsive nonlinearity, the reduced system produces, for the same
norm, narrower solitons. Further, Fig. \ref{FIG5} displays dependences of
the chemical potential and amplitude of the MMs on $N$, as found in the
reduced system, in comparison with the same dependences produced by the full
system. The absence of the repulsive terms makes the soliton's amplitude
higher, and the chemical potential more negative, for the same $N$.

\begin{figure}[tbp]
\centering
\begin{tabular}{lll}
(a) & (b) &  \\
\includegraphics[width=0.24\textwidth]{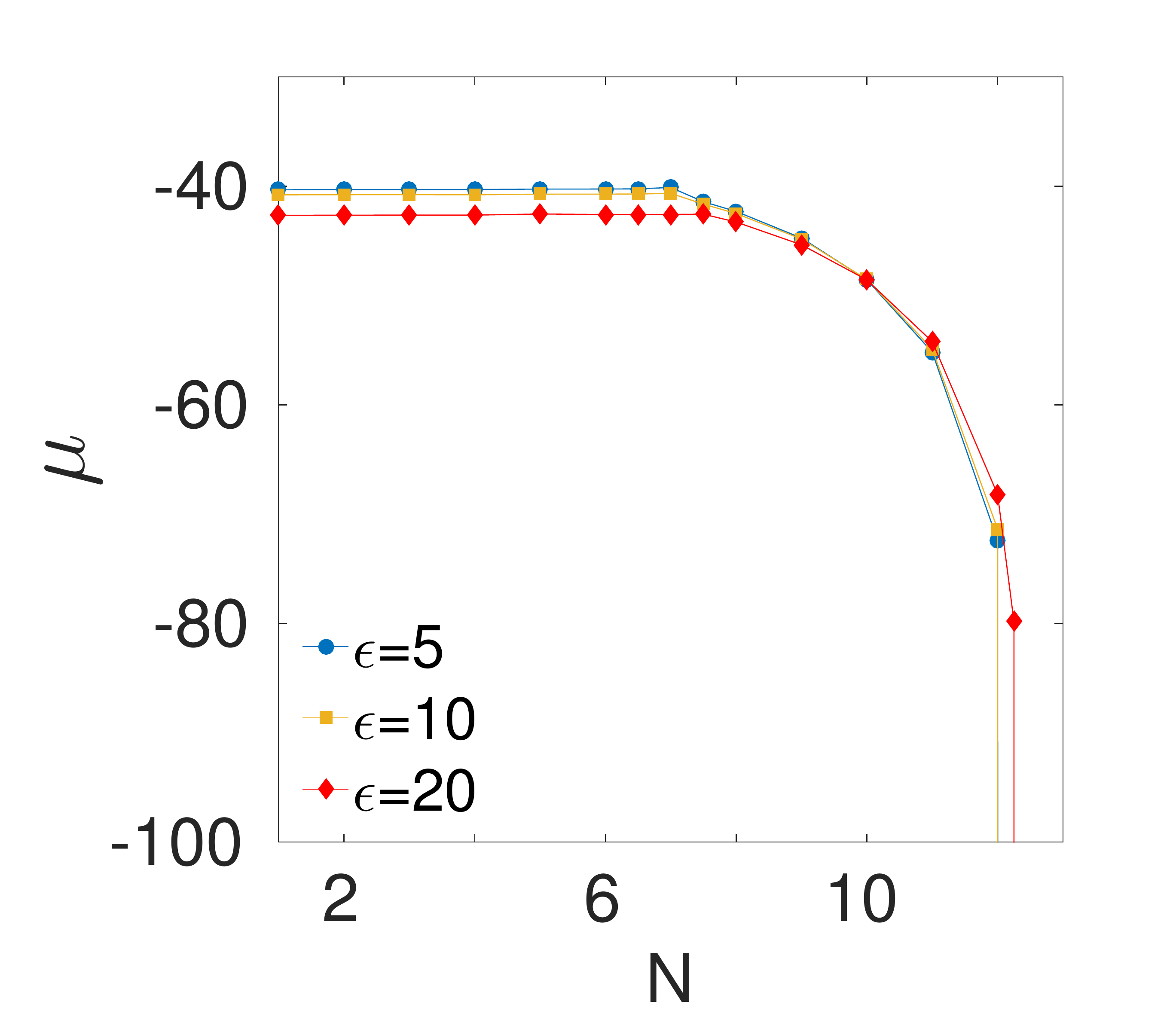} & %
\includegraphics[width=0.24\textwidth]{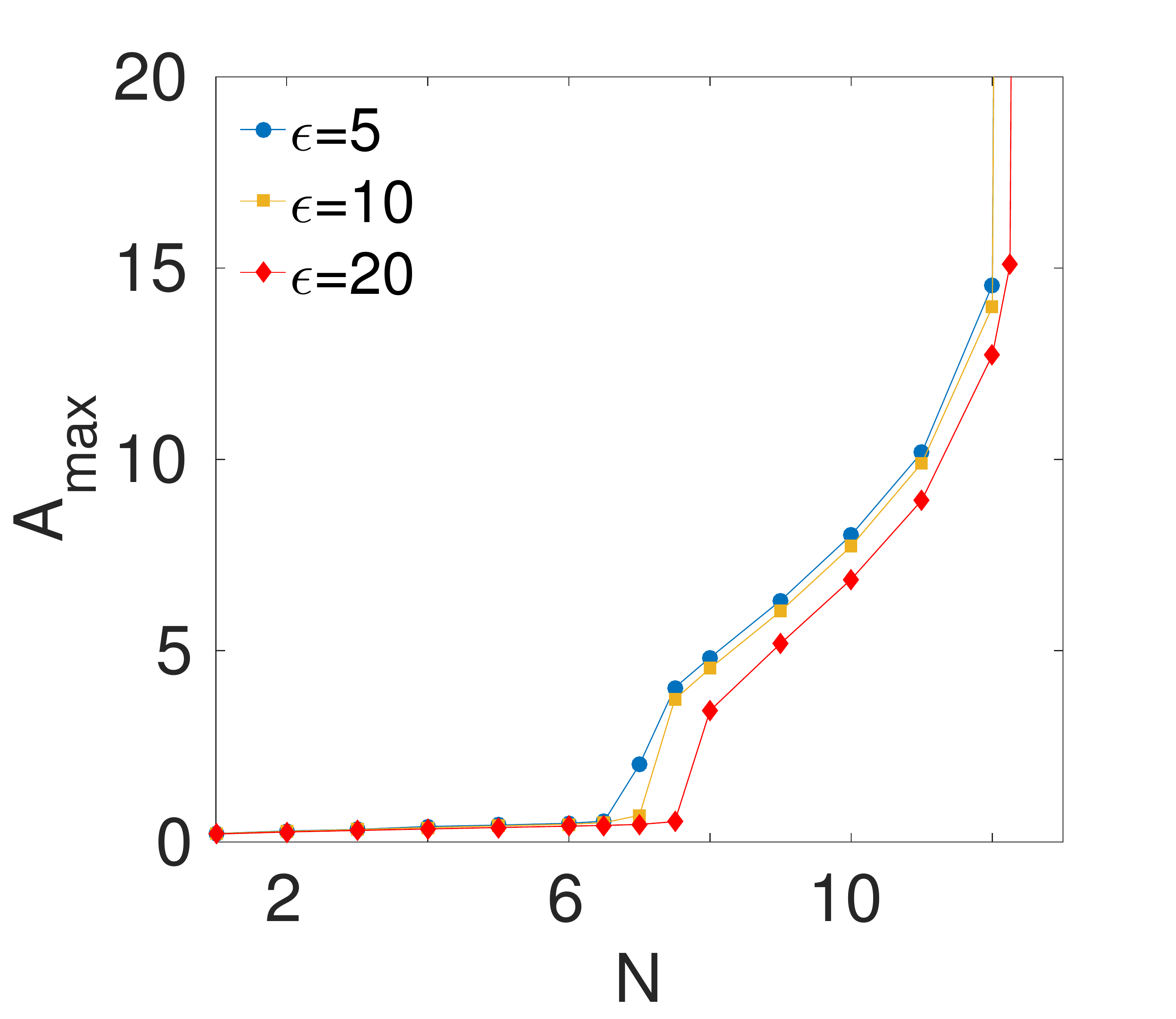} &  \\
&  &
\end{tabular}%
\caption{Chemical potential, $\protect\mu $, and maximum amplitude, $A_{\max
}$, as a function of $N$ for different values of the Zeeman-splitting
strength, $\protect\varepsilon $. Other fixed parameters are the same as in
Fig. \protect\ref{FIG1}. }
\label{FIG6}
\end{figure}

Finally, taking into account the Zeeman splitting, represented by
coefficient $\varepsilon $ in Eqs. (\ref{system}), Fig. \ref{FIG6} shows its
effect on the MM's chemical potential and amplitude. While the effect is not
dramatic, it somewhat additionally stabilizes the MM, making its chemical
potential more negative with the increase of $\varepsilon $. In addition to
that, Fig. \ref{FIG6}(b) shows that the increase of $\varepsilon $ increases
the threshold value $N_{\min }$ above which the MMs exist.

\section{Moving solitons and collisions between them}

\label{SIV}

\subsection{Soliton mobility}

\begin{figure}[tbp]
\centering
\begin{tabular}{lll}
(a) & (b) &  \\
\includegraphics[width=0.24\textwidth]{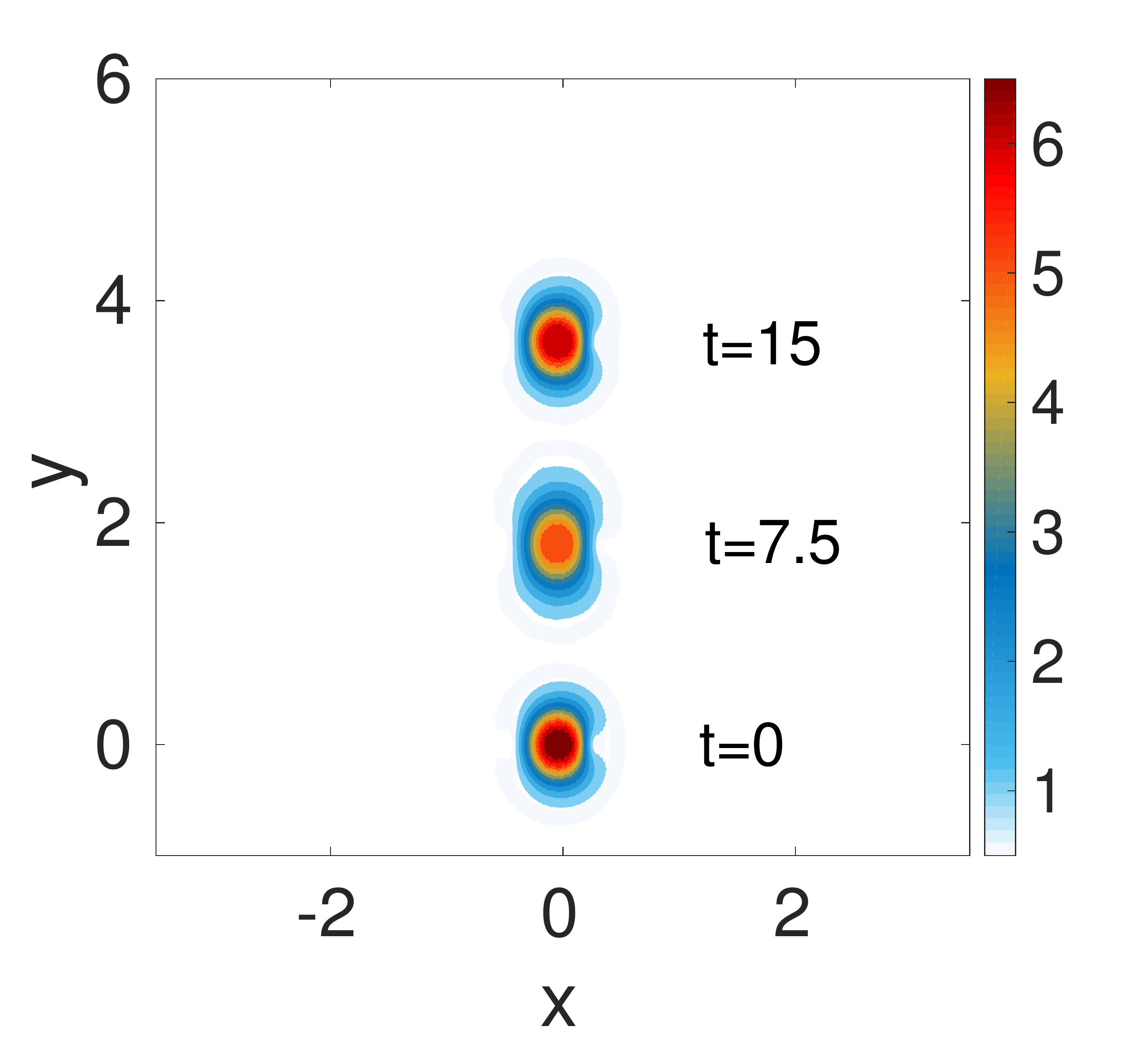} & %
\includegraphics[width=0.24\textwidth]{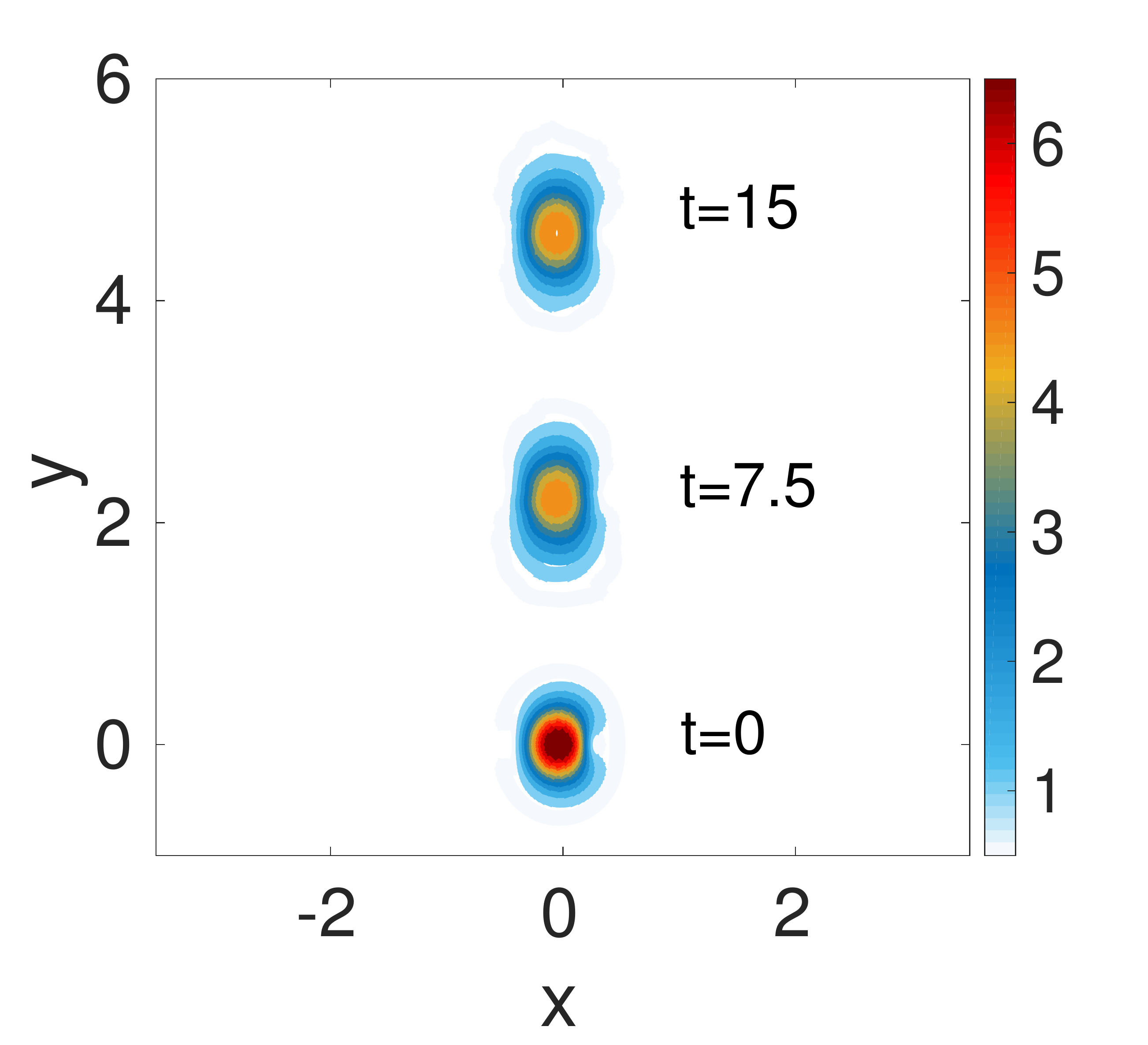} &  \\
&  &
\end{tabular}%
\caption{Contour plots of $|\Psi _{1}\left( x,y\right) |$ in the $\left(
x,y\right) $ plane for a mixed-mode soliton, suddenly set in motion by kicks
with $v_{y}=0.21$ (a) and $v_{y}=0.31$ (b). The evolution of the solitons is
displayed at three moments of time, as indicated in the panels. Fixed
parameters here are $\protect\gamma =1$, $\protect\varepsilon =0$, $\protect%
\lambda =8$, and $N=10$.}
\label{FIG7}
\end{figure}

\begin{figure}[tbp]
\centering
\begin{tabular}{lll}
(a) & (b) &  \\
\includegraphics[width=0.24\textwidth]{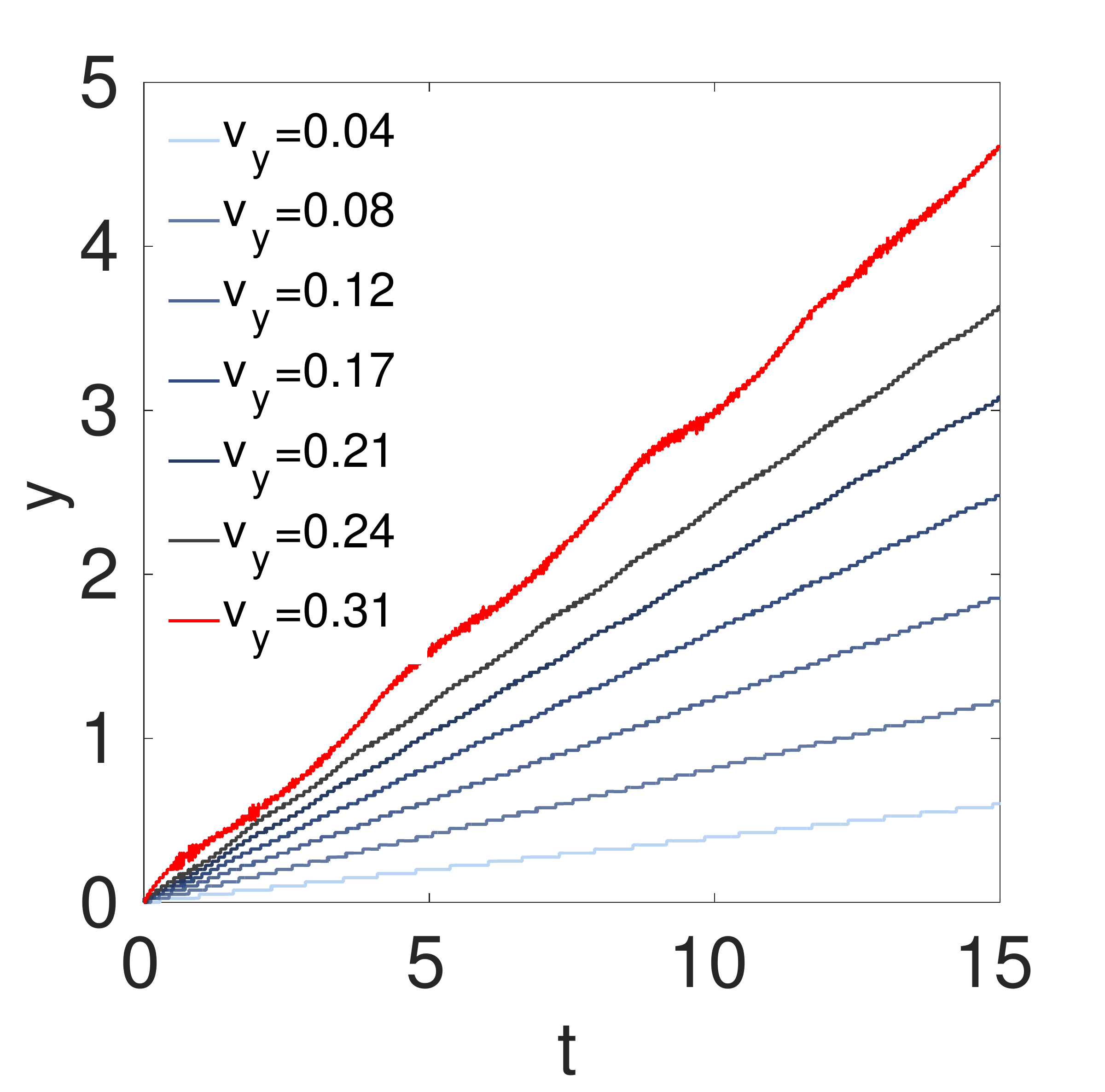} & %
\includegraphics[width=0.24\textwidth]{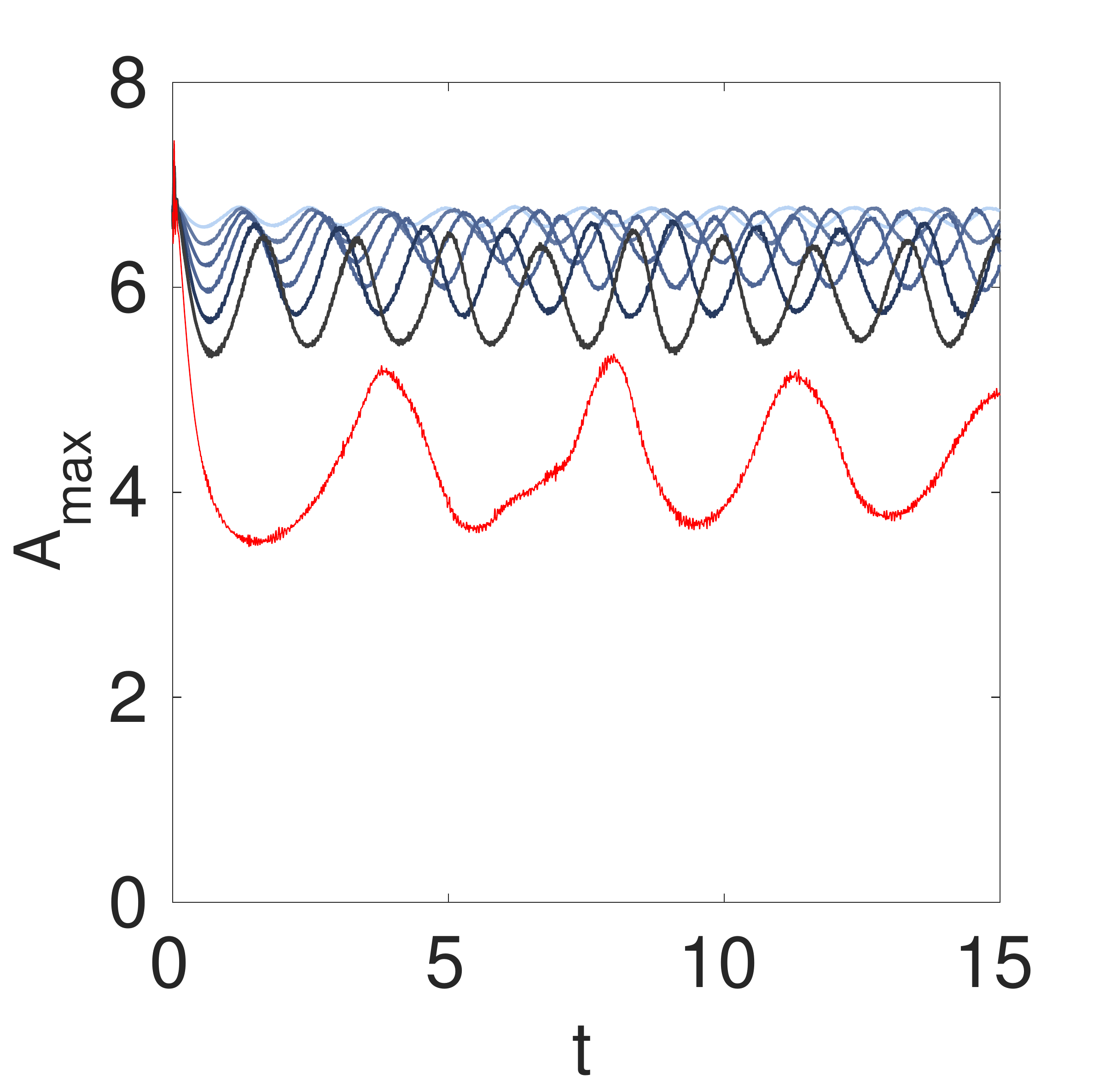} &  \\
&  &
\end{tabular}%
\caption{(a) The \ coordinate of the position of the maximum density
(center), and the amplitude of the kicked soliton, versus time. Fixed
parameters are the same as in Fig. \protect\ref{FIG7}.}
\label{FIG8}
\end{figure}

The SOC terms break the Galilean invariance of the underlying equations,
therefore generation of the moving solitons from quiescent ones is a
nontrivial problem for SOC systems \cite{Sakaguchi14}. To create such modes,
moving in the $y$ direction, in direct simulations, we apply a kick to the
originally quiescent MM soliton, multiplying both components of the spinor
wave function by an exponential factor, $\exp (iky)$. Figure \ref{FIG7}
shows two examples of the so generated spatiotemporal dynamics of the
soliton's component $|\Psi _{1}|$. Naturally, imparting larger momentum $k$
to the soliton gives rise to faster motion. Furthermore, Fig. \ref{FIG8}
shows that the solitons move with an oscillating component in their velocity
and amplitude. Solitons moving with higher velocities are found to be wider,
and they demonstrate larger oscillations. This feature is a precursor of the
destruction of solitons which are set in too fast motion, see below.

Alternatively, following work \cite{Sakaguchi14}, solutions for solitons
steadily moving at a given velocity $\mathbf{v}=(v_{x},v_{y})$ can be
obtained from Eq. (\ref{system}) rewritten in the reference frame moving
with velocity $\mathbf{v}$:%
\begin{eqnarray}
i\frac{{\partial {\Psi _{1}}}}{{\partial t}}-i\left( \mathbf{v}\cdot \nabla
\right) \Psi _{1} =-\frac{1}{2}{\nabla ^{2}}{\Psi _{1}}+{\left\vert {{\Psi
_{1}}}\right\vert ^{4/3}}{\Psi _{1}}-\gamma {\left\vert {{\Psi _{2}}}%
\right\vert ^{2}}{\Psi _{1}} +\lambda \left( {\frac{{\partial {\Psi _{2}}}}{{\partial x}}-i\frac{{%
\partial {\Psi _{2}}}}{{\partial y}}}\right) ,  \notag
\end{eqnarray}%
\begin{eqnarray}
i\frac{{\partial {\Psi _{2}}}}{{\partial t}}-i\left( \mathbf{v}\cdot \nabla
\right) \Psi _{2} =-\frac{1}{2}{\nabla ^{2}}{\Psi _{2}}+{\left\vert {{\Psi
_{2}}}\right\vert ^{4/3}}{\Psi _{2}}-\gamma {\left\vert {{\Psi _{1}}}%
\right\vert ^{2}}{\Psi _{2}} 
-\lambda \left( {\frac{{\partial {\Psi _{1}}}}{{\partial x}}+i\frac{{%
\partial {\Psi _{1}}}}{{\partial y}}}\right) .  \label{moving-reference}
\end{eqnarray}%
Stable stationary MM solitons have been produced by the numerical solution
of Eq. (\ref{moving-reference}), for the velocity vector $\mathbf{v}%
=(0,v_{y})$ (moving solitons with $v_{x}\neq 0$ could not be found).
Obviously, such solitons are moving ones, in terms of the laboratory
reference frame. Note that the stationary version of Eqs. (\ref%
{moving-reference}), produced by the substitution of ansatz (\ref{U}),
features the same symmetry which is evident in Eqs. (\ref{UU}) \cite{BLi}:
\begin{equation}
U_{1,2}(x,y)=U_{2,1}\left( -x,y\right). 
\label{symm}
\end{equation}%

\begin{figure}[tbp]
\centering
\resizebox{0.4\textwidth}{!}{
\includegraphics{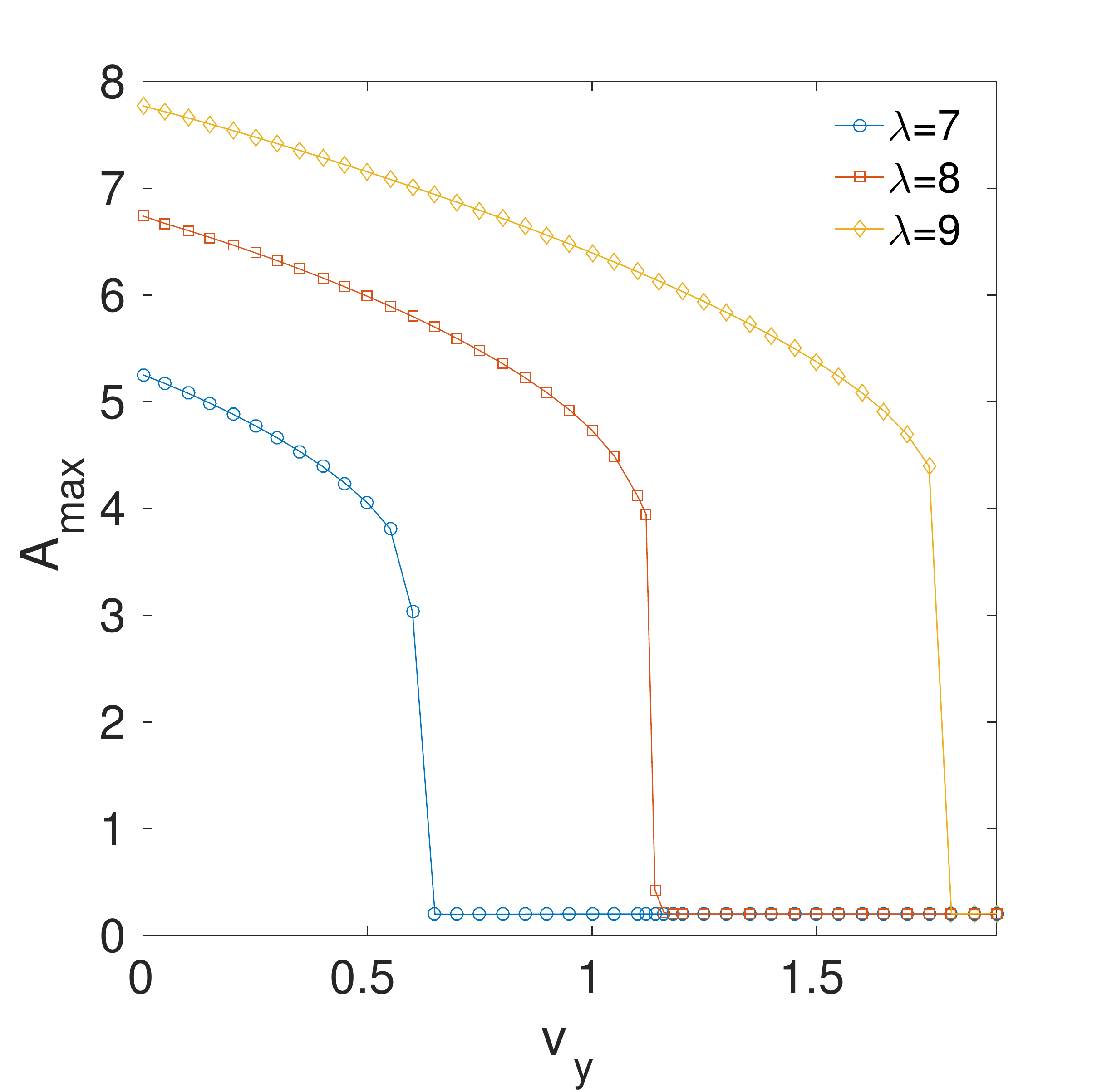}
}
\caption{The amplitude of moving solitons, produced as stationary solutions
of Eqs. (\protect\ref{moving-reference}), as a function of $v_{y}$. The
moving solitons cease to exist when the amplitude falls to zero. Parameters
are the same as in Fig. \protect\ref{FIG7}.}
\label{FIG9}
\end{figure}

Figure \ref{FIG9} shows the amplitude of the steadily moving MM, produced by
the solution of Eq. (\ref{moving-reference}), as a function of $v_{y}$, at
three different values of the SOC strength, $\lambda $. It is seen that the
amplitude decreases with the growth of the velocity, but increases with the
growth of $\lambda $. At the critical value of $v_{y}$, the amplitude
abruptly vanishes, which implies nonexistence of MMs moving with velocities
exceeding the critical value. In the binary-BEC system, SOC also prevents
the existence of MM\ solitons at velocities exceeding a critical value \cite%
{Sakaguchi14}, but the vanishing of the amplitude with the increase of $%
v_{y} $ is smoother than in the present case. On the other hand, the
critical velocity increases with $\lambda $, showing that the solitons'
stability enhances for sharper solitons.

\begin{figure}[tbp]
\centering
\begin{tabular}{lll}
(a) & (b) &  \\
\includegraphics[width=0.24\textwidth]{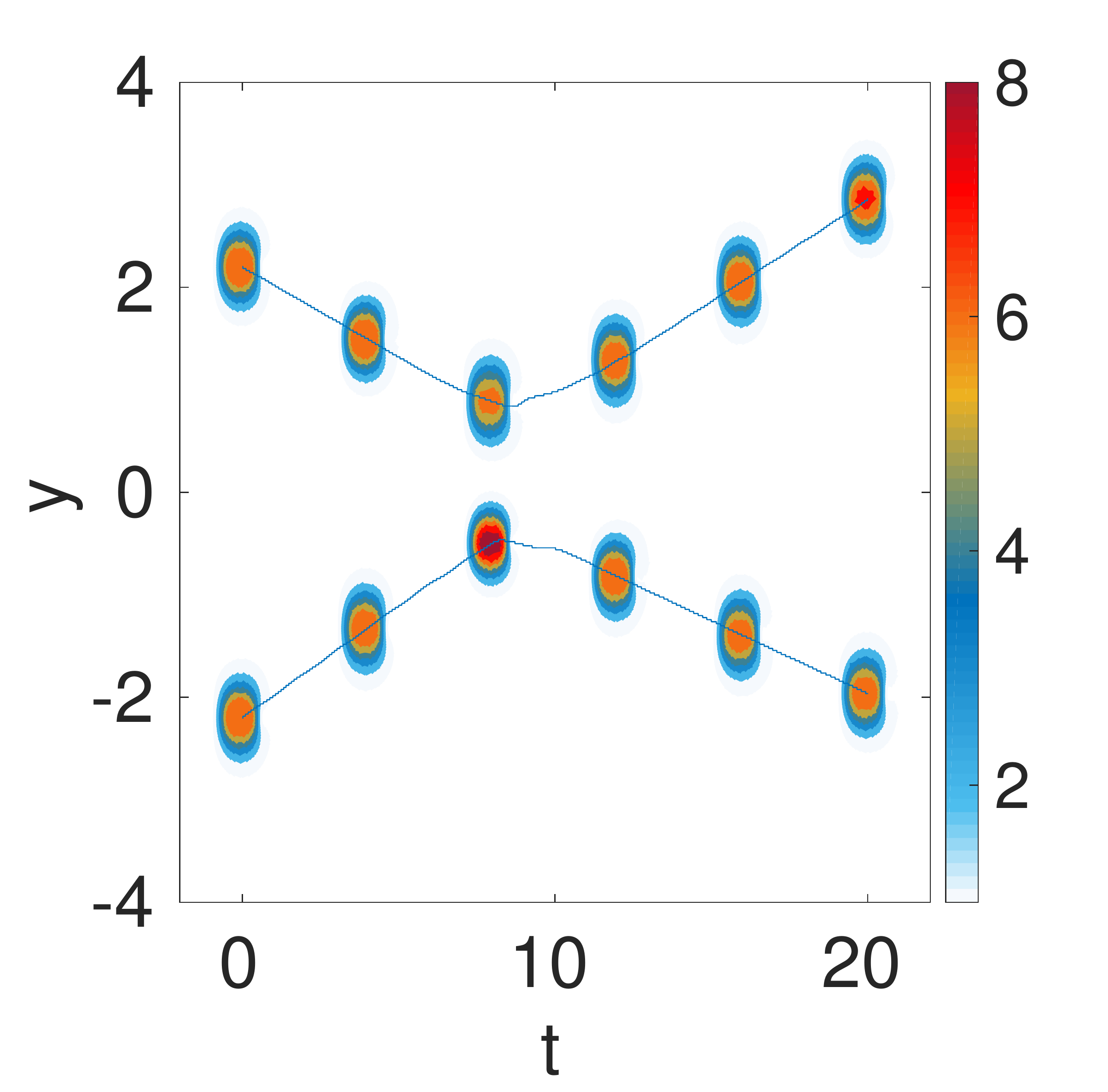} & %
\includegraphics[width=0.24\textwidth]{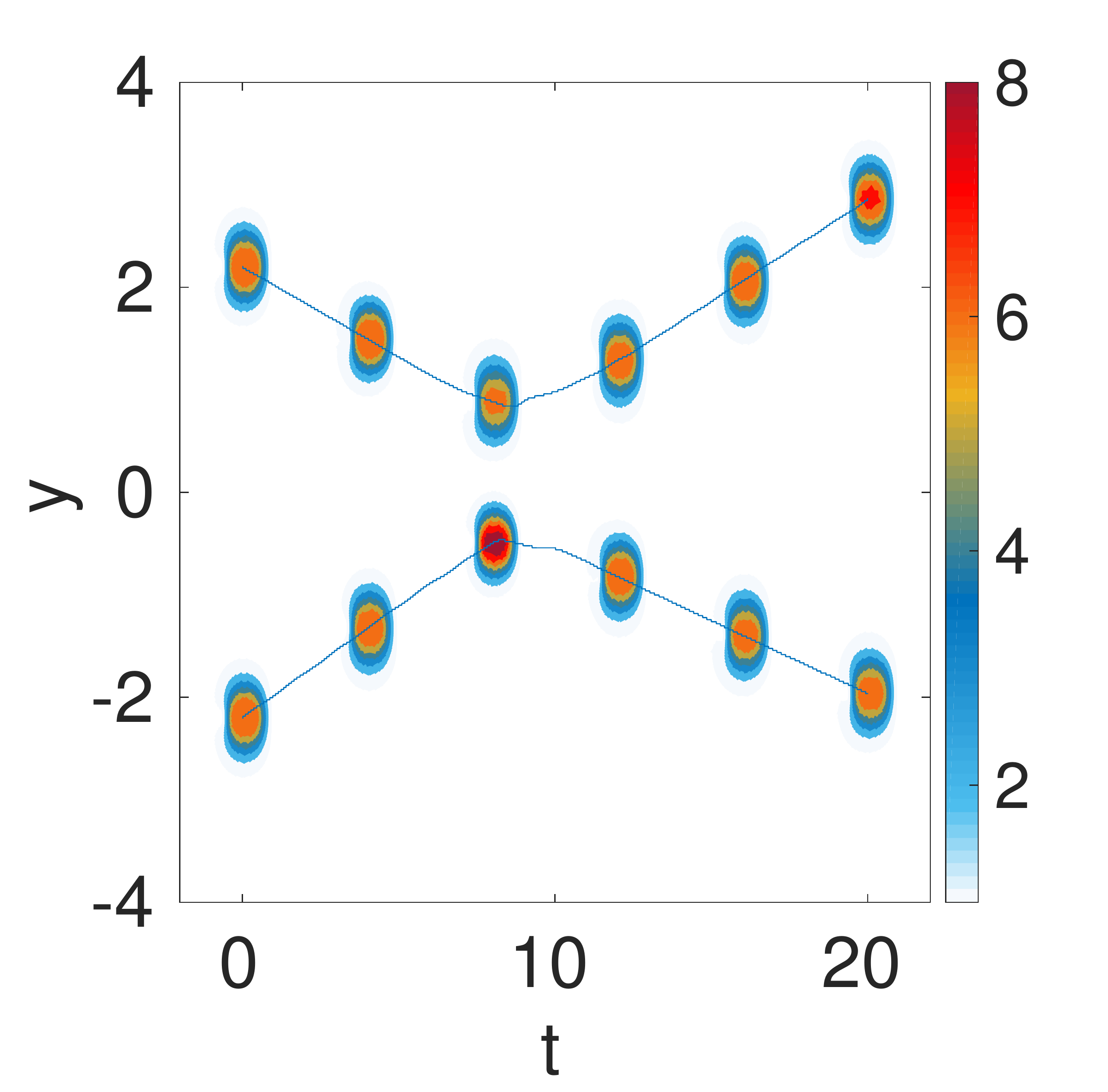} &  \\
&  &  \\
(c) & (d) &  \\
\includegraphics[width=0.24\textwidth]{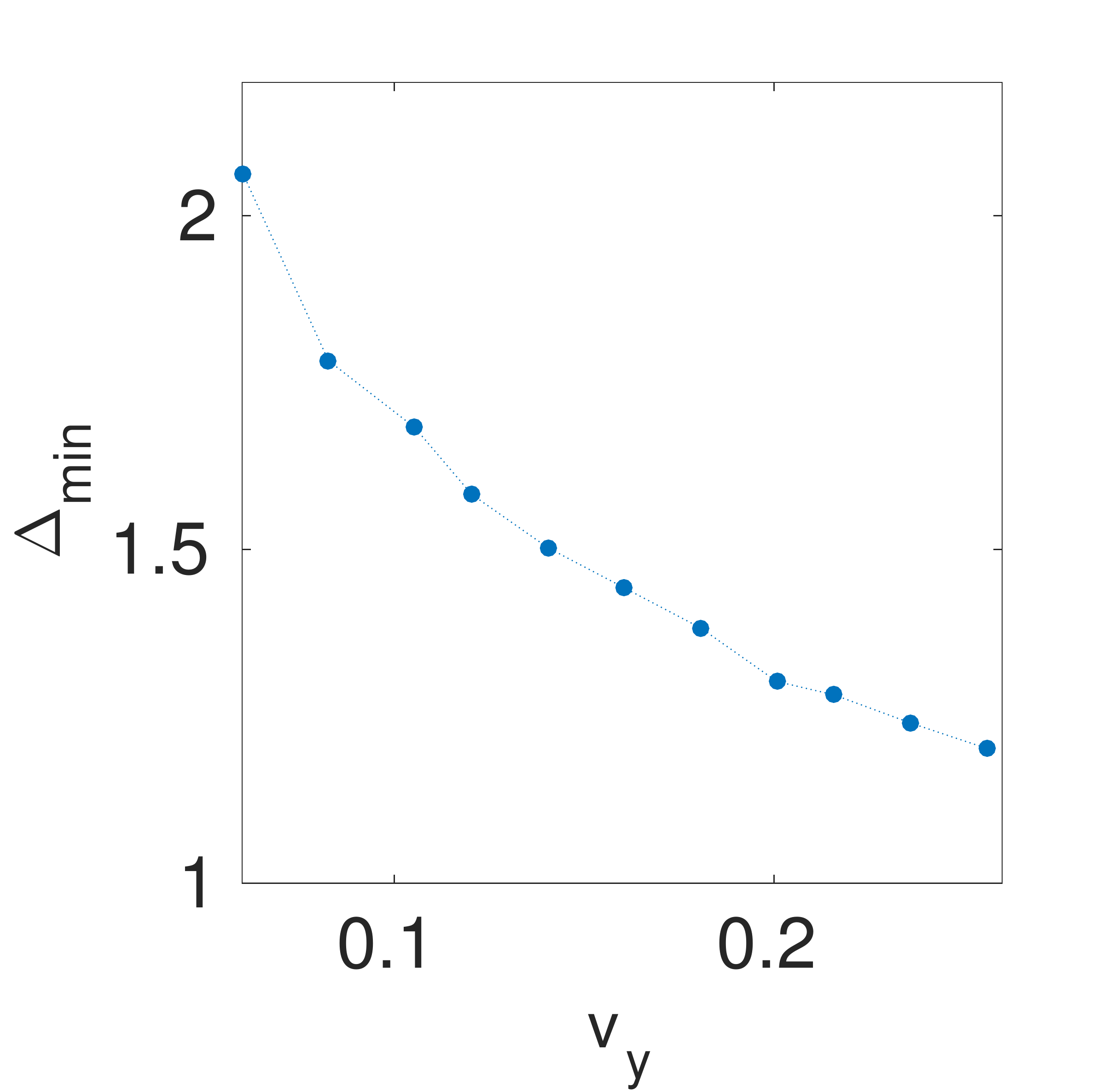} & %
\includegraphics[width=0.24\textwidth]{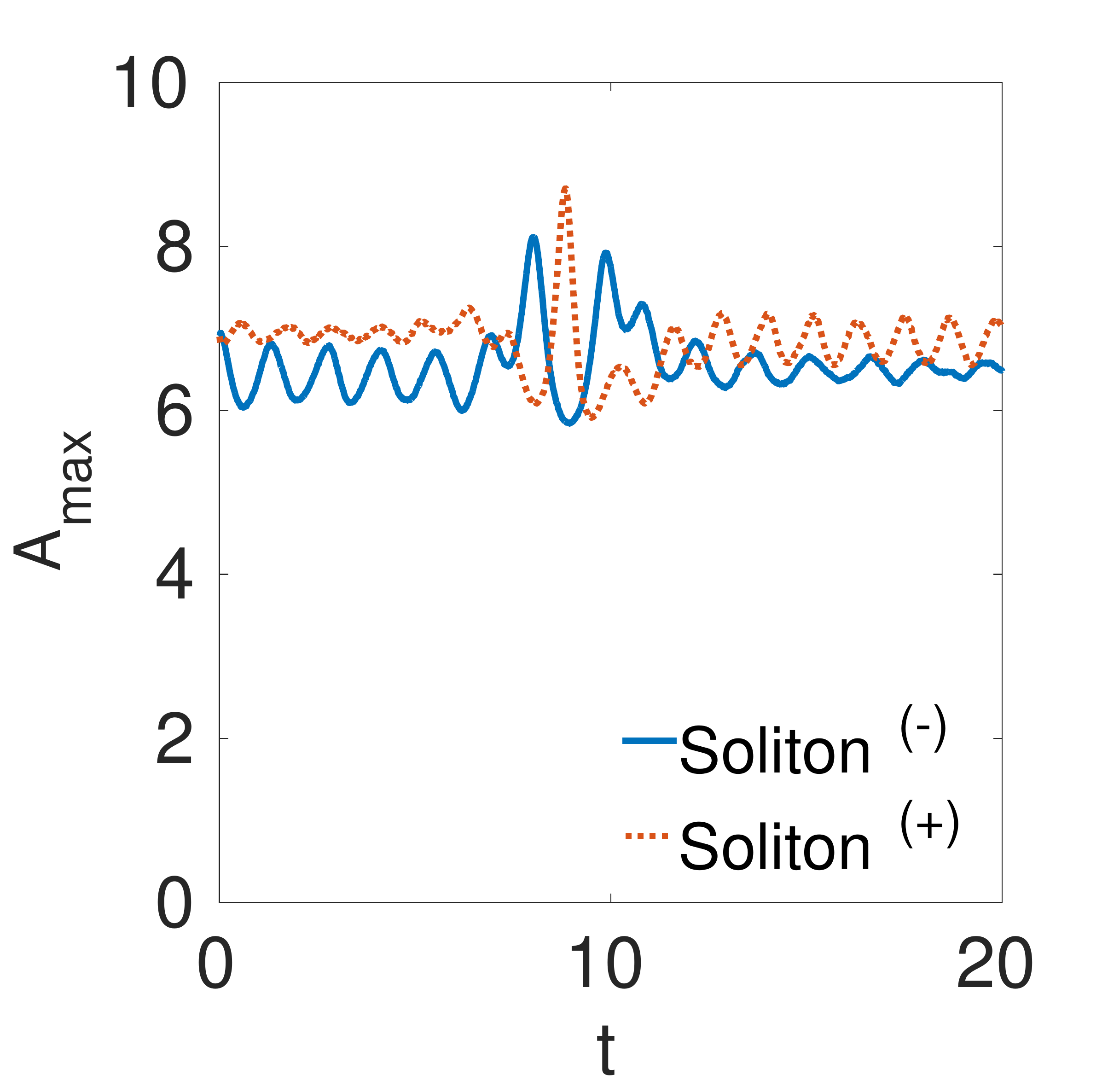} &  \\
&  &
\end{tabular}%
\caption{A typical example of elastic collisions between identical solitons
moving in opposite directions is shown by means of contour plots of both
component, $|\Psi _{1}|$ (a) and $|\Psi _{2}|$ (b), in the plane of $(x,y)$,
at several moments of time, $t$, as a function of the $y$-position of the
maximum amplitude and time. (c) The minimum separation between the bouncing
solitons, $\Delta _{\min }$, as a function of the initial velocity, $\pm
v_{y}$. The fixed parameters are $v_{y}=0.2$ and $N=20$, while the others
are the same as in Fig. \protect\ref{FIG7}. (d) Amplitudes of the colliding
solitons as functions of time, with \textquotedblleft Soliton$^{(-)}$" and
\textquotedblleft Soliton$^{(+)}$" referring to the solitons originally
placed at $y<0$ and $y>0$, respectively. }
\label{FIG10}
\end{figure}

\begin{figure}[tbp]
\centering
\begin{tabular}{lll}
(a) & (b) &  \\
\includegraphics[width=0.24\textwidth]{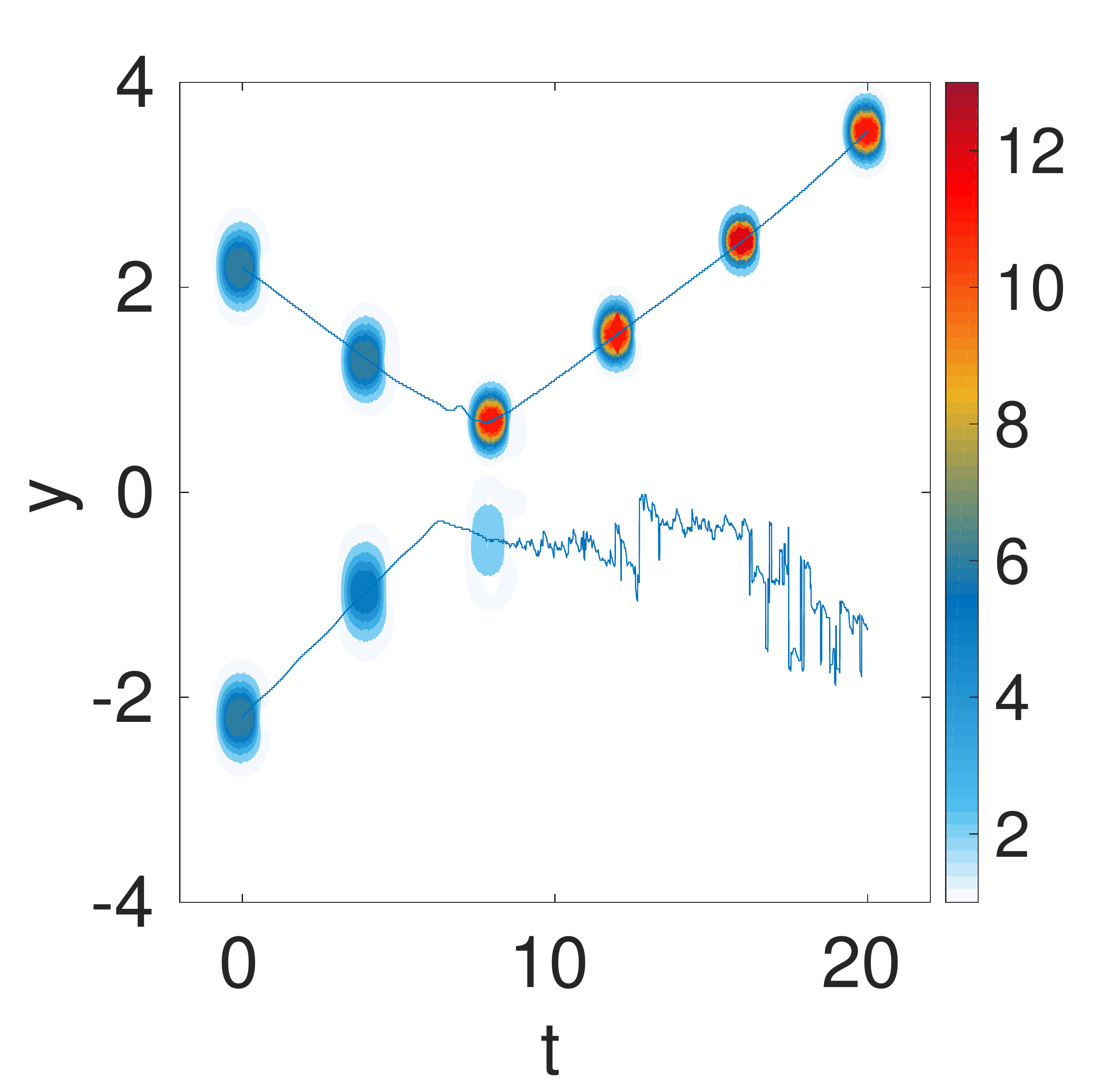} & %
\includegraphics[width=0.24\textwidth]{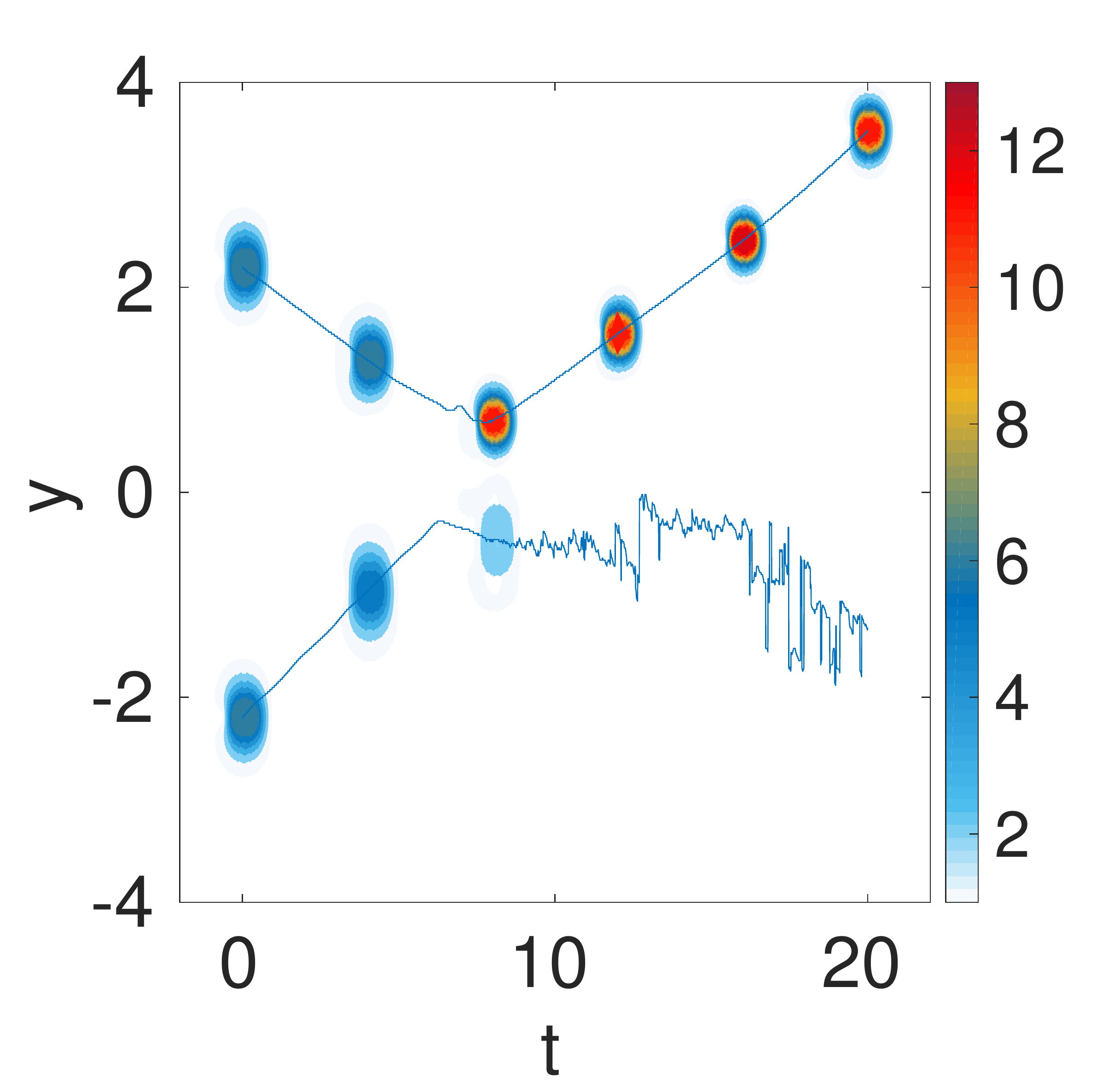} &  \\
&  &  \\
&  &
\end{tabular}
\begin{tabular}{lll}
& (c) &  \\
& \includegraphics[width=0.24\textwidth]{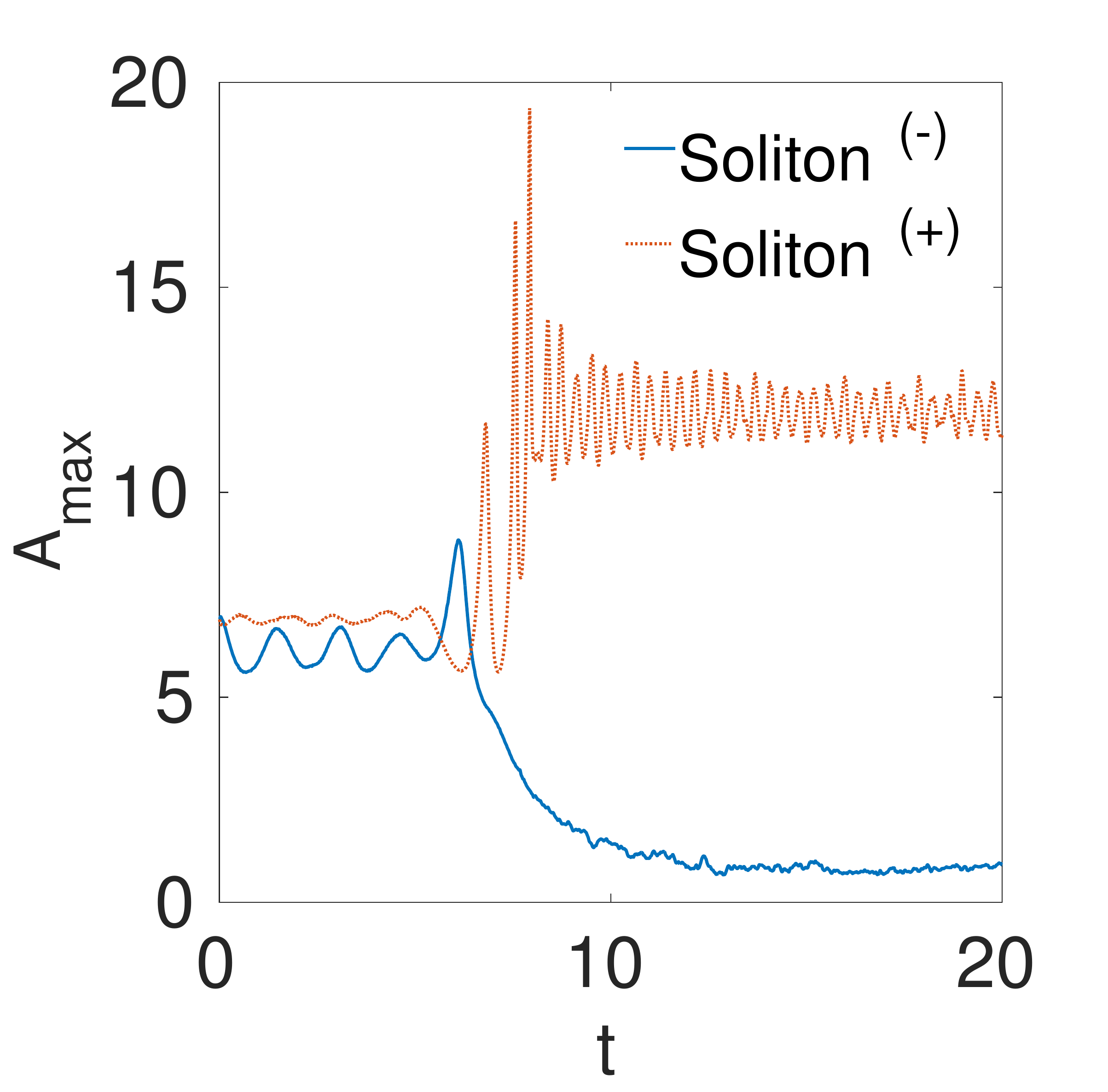} &  \\
&  &
\end{tabular}%
\caption{An inelastic collision leading to destruction of one soliton.
Panels (a), (b), and (c) have the same meaning as (a), (b), and (d),
respectively, in Fig. \protect\ref{FIG10}. The parameters are $v_{y}=0.27$, $%
N=20$, the others being the same as in Fig. \protect\ref{FIG7}.}
\label{FIG11}
\end{figure}

\subsection{Collisions between counterpropagating solitons}

\label{SV}

The availability of the moving solitons suggests to consider collisions
between them. Solutions were initiated by taking a pair of solitons
separated by distance $d$, kicked by factors $\exp (\pm i\mathbf{k}\cdot
\mathbf{r})$. A typical picture, displayed in Fig. \ref{FIG10}, demonstrates
that the colliding solitons bounce back from each other, after reaching a
minimal distance, $\Delta _{\min }$. This distance is shown in Fig. \ref%
{FIG10}(c) as a function of the initial velocity, $\pm v_{y}$. The decrease
of $\Delta _{\min }$ with the increase of $v_{y}$ is a natural result.
Indeed, if the effective potential of the repulsion between the 2D solitons
depends on the distance between them, $\Delta $, in the usual exponential
form, $U_{\mathrm{rep}}\simeq U_{0}\exp \left( -\Delta /\Delta _{0}\right) $%
, with positive $U_{0}$ and $\Delta _{0}$ \cite{potential}, the balance of
the repulsive potential and the solitons' kinetic energy (provided that it
may be estimated as in the Galilean-invariant systems, $E_{\mathrm{kin}}\sim
v_{y}^{2}$, predicts $\Delta _{\min }\sim \ln \left( 1/v_{y}\right) $, for
relatively small $v_{y}$.

The dependence of amplitudes of the colliding solitons on time, displayed in
Fig. \ref{FIG10}(d), shows that the initial kicks excite internal
oscillations in the solitons [cf. Fig. \ref{FIG8})], as they break symmetry (%
\ref{symm}) of the stationary MMs, and the kick of opposite signs, $\pm k$,
break it differently. Further, Fig. \ref{FIG10}(c) demonstrates that the
collision leads to the switch of inner oscillations between the two solitons.

In addition to elastic collisions shown in Fig. \ref{FIG10}, the simulations
reveal another outcome, in the form of partial destruction, as shown in Fig. %
\ref{FIG11}: one soliton bounces back in a strongly excited state, while the
other one (the soliton which featured conspicuous intrinsic excitation as
the result of the initial kick) suffers destruction. Note that the amplitude
(hence also the norm) of the surviving soliton essentially exceed their
initial values, as it absorbs a large share of the norm from the destroyed
soliton. For this reason, such an outcome of the collision may be
categorized as partial merger of the colliding solitons.

Detailed study of collisions demonstrates additional outcomes (which can be
observed, in particular, by varying the norm of the solitons), such as
quasi-elastic passage of solitons through each other, excitation of strong
inner oscillations, and destruction of both solitons. Complete results for
collisions will be presented elsewhere.

\section{Conclusion}

\label{SVI}

The objective of this work is to construct stable self-trapped
vortex-soliton complexes in the 2D model of the Fermi gas with two
components, representing spin-up and down-polarized atomic states, linearly
interacting through by the SOC (spin-orbit-coupling) terms of the Rashba
type and contact nonlinear attraction, competing with the effective Pauli
self-repulsion in each component, predicted by the density-functional theory
for the binary Fermi gas. In addition to its physical significance, the
model is interesting as it makes it possible to study the interplay of the
linear SOC effect with competing self-repulsive and cross-attractive terms,
which feature different powers of the nonlinearity. As a limit case,
corresponding to the domination of the inter-component attraction, the
system without the intrinsic repulsion was also considered, which
corresponds to a binary bosonic gas as well. The systematic numerical
analysis has produced a parameter region populated by stable two-component
solitons of the MM (mixed-mode) type, while all the SV (semi-vortex)
solitons are found to be unstable. The largest velocity up to which the
solitons may travel in this system with the broken Galilean invariance has
been identified too. Finally, we have briefly addressed collisions between
solitons moving in opposite directions, demonstrating various outcomes of
the collisions, such as quasi-elastic rebound of the solitons from each
other, and destruction of one of them.

The analysis reported in this work can be extended by including a
combination of the Rashba and Dresselhaus types of SOC (cf. the analysis
performed in Ref. \cite{Sherman} for 2D solitons in binary BEC). Another
interesting possibility is to consider the limit case of \textquotedblleft
heavy atoms", i.e., negligible kinetic-energy terms in Eq. (\ref{system}).
As shown in terms of the binary bosonic gas under the action of SOC, in this
case the dispersion law (\ref{eps}) is replaced by one which features a
\textit{gap}, $\mu =\pm \sqrt{\varepsilon ^{2}+\lambda ^{2}k^{2}}$, that can
be populated by \textit{gap solitons} \cite{HS}. The consideration of the
respective model for the binary Fermi gas will be reported elsewhere.
Lastly, a challenging possibility is to extend the work for 3D settings, cf.
the prediction of metastable solitons supported by SOC in the 3D bosonic gas
\cite{HP}.

\section*{Acknowledgments}

PD acknowledges partial financial support from DIUFRO project under grant
DI18-0066 and CMCC of the Universidad de La Frontera. DL acknowledge partial
financial support from Centers of excellence with BASAL/CONICYT financing,
Grant FB0807, CEDENNA and CONICYT-ANILLO ACT 1410. PD and DL acknowledges
financial support form FONDECYT 1180905. The work of BAM is supported, in
part, by the joint program in physics between NSF and Binational (US-Israel)
Science Foundation through project No. 2015616, and by the Israel Science
Foundation, through grant No. 1287/17. The authors appreciates a support
provided by the PAI-CONICYT program (Chile), grant No. 80160086, and
hospitality of Instituto de Alta Investigaci\'{o}n, at Universidad de Tarapac%
\'{a} (Arica, Chile).

\end{document}